\documentclass[prl,twocolumn,preprintnumbers,superscriptaddress,amsmath,amssymb]{revtex4}
\usepackage{graphicx}
\usepackage{subfigure}
\usepackage{mathrsfs}
\usepackage{amsfonts}
\usepackage{amsmath}
\usepackage{leftidx}
\usepackage{color}
\usepackage[colorlinks,linkcolor=blue,citecolor=blue]{hyperref}

\begin{document}
\title{Zeno Hall Effect}
\author{Zongping Gong}
\affiliation{Department of Physics, University of Tokyo, 7-3-1 Hongo, Bunkyo-ku, Tokyo 113-0033, Japan}
\author{Sho Higashikawa}
\affiliation{Department of Physics, University of Tokyo, 7-3-1 Hongo, Bunkyo-ku, Tokyo 113-0033, Japan}
\author{Masahito Ueda}
\affiliation{Department of Physics, University of Tokyo, 7-3-1 Hongo, Bunkyo-ku, Tokyo 113-0033, Japan}
\affiliation{RIKEN Center for Emergent Matter Science (CEMS), Wako, Saitama 351-0198, Japan}
\date{\today}

\begin{abstract}
We show that the quantum Zeno effect gives rise to the Hall effect by tailoring the Hilbert space of a two-dimensional lattice system into a single Bloch band with a nontrivial Berry curvature. Consequently, a wave packet undergoes transverse motion in response to a potential gradient -- a phenomenon we call the \emph{Zeno Hall effect} to highlight its quantum Zeno origin. 
The Zeno Hall effect leads to retroreflection at the edge of the system due to an interplay between the band flatness and the nontrivial Berry curvature. We propose an experimental implementation of this effect with ultracold atoms in an optical lattice. 
\end{abstract}
\maketitle

\emph{Introduction.---} The state-of-the-art experimental techniques in atomic, molecular and optical physics have made it possible not only to engineer the Hamiltonian of a quantum system, but also to control its interaction with the environment \cite{Cirac2011,Blatt2011,Blatt2013,Wineland2013,Siddiqi2016,Marquardt2017}. Here controlled dissipation can serve as a resource for quantum coherence and entanglement \cite{Zoller1996,Buchler2008}, with versatile applications to quantum-state preparation \cite{Diehl2011,Diehl2012,Diehl2015}, quantum computation \cite{Verstraete2009} and quantum simulation \cite{Buchler2010,Zoller2014}.

The experimental progress in turn has stimulated theoretical studies in open quantum systems \cite{Diehl2008,Carusotto2009,Hartmann2010,Koch2010}, such as the Hall effects in the presence of dissipation \cite{Avron2012,Carusotto2012,Carusotto2014,Jiang2016}. It has been shown that the quantization of the transverse conductivity in the integer quantum Hall regime is robust against dissipation \cite{Avron2012,Jiang2016}, while a nontrivial influence of dissipation emerges for the fractional quantum Hall effect \cite{Carusotto2012} and the anomalous Hall effect \cite{Carusotto2014}.

\begin{figure}
\begin{center}
        \includegraphics[width=8cm, clip]{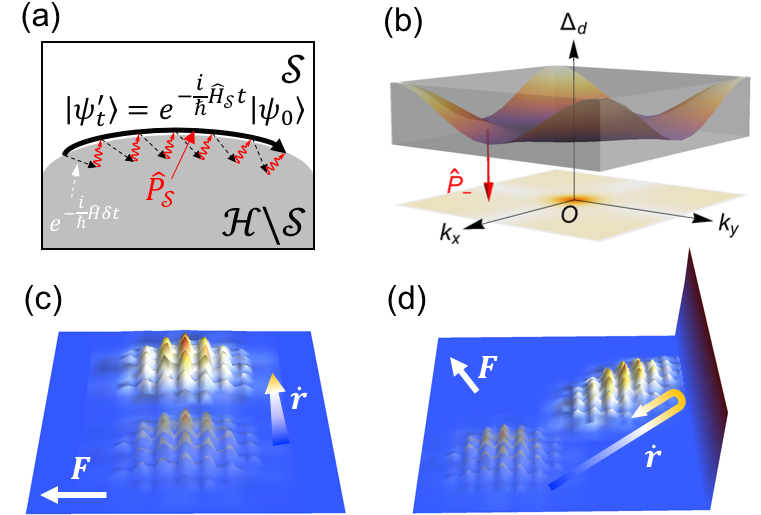}\\
      \end{center}
   \caption{(color online). (a) When a Hilbert space $\mathcal{H}$ is dissipatively constrained to a subspace $\mathcal{S}$, 
   the dynamics is governed by the projected Hamiltonian $\hat H_{\mathcal{S}}=\hat P_{\mathcal{S}}\hat H\hat P_{\mathcal{S}}$, where $\hat H$ and $\hat P_{\mathcal{S}}$ are the bare Hamiltonian and the projection operator onto $\mathcal{S}$, respectively. (b) In a strongly dissipative lattice system in two dimensions, the occupation of the upper band is prohibited by a large damping gap $\Delta_d$ due to the quantum Zeno effect, and the system is constrained to the lower band with a nonzero Berry curvature. Consequently, a 
   wave packet undergoes transverse motion in response to a potential gradient $\boldsymbol{F}$ (c), and will be retroreflected at a boundary (d).}
     \label{fig1}
\end{figure}

In this Letter, we point out yet another Hall effect in open quantum systems --- the Hall effect \emph{due to} dissipation. This differs fundamentally from the previous works in that the Hall effect originates from the interaction with the environment instead of the bare Hamiltonian $\hat H$ of the system. Our idea is based on two key ingredients: (i) to use dissipation to tailor the accessible Hilbert space $\mathcal{S}$ and hence to change the effective Hamiltonian (see Fig.~\ref{fig1} (a)) \cite{Knight2000,Smerzi2014,Reichel2015}; (ii) the noncommutativity of position operators in a constrained Bloch band with a nontrivial Berry curvature \cite{Niu2010}.  (i) exploits the quantum Zeno (QZ) effect \cite{Sudarshan1977,Facchi2000,Facchi2002}, which is well studied in the context of quantum measurement \cite{Mekhov2015,Mekhov2016b} and occurs also for strong dissipation \cite{Cirac2008,Ott2013,Rey2014} as a continuous limit of repeated measurements \cite{Knight2000}. (ii) shares the same physics behind the anomalous Hall effect \cite{Nagaosa2010}. As schematically illustrated in Figs.~\ref{fig1} (b) and (c), if the entire Hilbert space is tailored into a single Bloch band with a nonzero Berry curvature, a wave packet undergoes transverse motion even if $\hat H$ has no kinetics term. We call such a phenomenon the Zeno Hall effect (ZHE). Our scheme is readily applicable to create a \emph{flat band} with a tunable Berry curvature, and thus provides an ideal platform to explore quantum many-body physics \cite{Tasaki2003,Katsura2010,Altman2010}. Surprisingly, the wave-packet dynamics in such a flat band is found to be \emph{retroreflective} (see Fig.~\ref{fig1} (d)). In light of the recent development of reservoir engineering \cite{Cirac2011,Blatt2011,Blatt2013,Wineland2013,Siddiqi2016,Marquardt2017} and artificial gauge fields \cite{Spielman2011,Dalibard2011,Ketterle2013,Bloch2013,Goldman2014,Bloch2015,Pan2016}, 
we expect that the ZHE can be implemented with, e.g., ultracold atoms. 

\emph{Tailoring the Hilbert space into a Bloch band.---} As a minimal setup, we consider spin-$1/2$ free fermions or bosons in a two-dimensional  $N\times N$ ($N\gg1$) square lattice with the lattice constant $a$. We assume that the intrinsic kinetics is completely quenched. Within the tight-binding and Born-Markov approximations \cite{Breuer2002}, the dissipative dynamics of the system can be modeled by the Lindblad equation \cite{Daley2014b,Diehl2016}
\begin{equation}
\dot{\hat\rho}=\Gamma\sum_{\boldsymbol{r}}\left(\hat L_{\boldsymbol{r}}\hat\rho\hat L^\dag_{\boldsymbol{r}}-\frac{1}{2}\{\hat L^\dag_{\boldsymbol{r}}\hat L_{\boldsymbol{r}},\hat\rho\}\right),
\label{LE}
\end{equation}
where $\hat\rho$ is the density operator of the system, $\boldsymbol{r}=(m,n)a$ ($m,n=1,2,...,N$) denotes a lattice site, and $\Gamma$ is the damping rate. We assume that the dissipation is caused by  one-body loss and respects the translational symmetry, so that $\hat L_{\boldsymbol{r}}=\sum_{\boldsymbol{r}',\sigma}l_{\boldsymbol{r}'-\boldsymbol{r},\sigma}\hat c_{\boldsymbol{r}'\sigma}$, where $\hat c_{\boldsymbol{r}'\sigma}$ is the annihilation operator of a particle with spin $\sigma$ ($\uparrow$ or $\downarrow$) at site $\boldsymbol{r}'$ and $l_{\boldsymbol{r}'-\boldsymbol{r},\sigma}$'s are $c$-numbers. 

Since there are $N^2$ $\hat L_{\boldsymbol{r}}$'s while the dimension of the single-particle Hilbert space is $2N^2$, the dimension of the single-particle decoherence-free subspace \cite{Lidar1998} is at least $N^2$. Furthermore, due to the translational invariance, if a wave function $\psi(\boldsymbol{r})$ is decoherence-free, so will be $\psi(\boldsymbol{r}+\boldsymbol{a}_\mu)$, where $\boldsymbol{a}_\mu=a\boldsymbol{e}_\mu$ ($\mu=x,y$), reflecting a band structure. We can explicitly show that Eq.~(\ref{LE}) is equivalent to
\begin{equation}
\dot{\hat\rho}=\Gamma\sum_{\boldsymbol{k}}\Delta_{\boldsymbol{k}}\left(\hat c_{\boldsymbol{k}+}\hat\rho\hat c^\dag_{\boldsymbol{k}+}-\frac{1}{2}\{\hat c^\dag_{\boldsymbol{k}+}\hat c_{\boldsymbol{k}+},\hat\rho\}\right),
\end{equation}
implying that such a decoherence-free band (see Fig.~\ref{fig1} (b)) is spanned by $|\boldsymbol{k}-\rangle=\hat c^\dag_{\boldsymbol{k}-}|{\rm vac}\rangle$, where $\hat c^\dag_{\boldsymbol{k}+}=\Delta^{-\frac{1}{2}}_{\boldsymbol{k}}(l_{\boldsymbol{k}\uparrow}\hat c^\dag_{\boldsymbol{k}\uparrow}+l_{\boldsymbol{k}\downarrow}\hat c^\dag_{\boldsymbol{k}\downarrow})$, $\hat c^\dag_{\boldsymbol{k}-}=\Delta^{-\frac{1}{2}}_{\boldsymbol{k}}(l_{\boldsymbol{k}\downarrow}\hat c^\dag_{\boldsymbol{k}\uparrow}- l_{\boldsymbol{k}\uparrow}\hat c^\dag_{\boldsymbol{k}\downarrow})$, $\Delta_{\boldsymbol{k}}=\sum_\sigma|l_{\boldsymbol{k}\sigma}|^2$ is the dimensionless damping gap rescaled by $\hbar\Gamma$ with $\boldsymbol{k}=\frac{2\pi}{Na}(m,n)$ ($m,n=-(N-1)/2,...,(N-1)/2$) being the crystal momentum, $l_{\boldsymbol{k}\sigma}=\sum_{\boldsymbol{r}}e^{i\boldsymbol{k}\cdot\boldsymbol{r}}l_{\boldsymbol{r}\sigma}$ and $\hat c_{\boldsymbol{k}\sigma}=N^{-1}\sum_{\boldsymbol{r}}e^{-i\boldsymbol{k}\cdot\boldsymbol{r}}\hat c_{\boldsymbol{r}\sigma}$ \cite{SMD}. Here we assume $\Delta_{\boldsymbol{k}}>0$ over the entire Brillouin zone. In the large-$\Gamma$ limit, the lower band is not only a decoherence-free subspace, but also a QZ subspace \cite{Facchi2002}. 

\emph{Wave-packet dynamics.---} Governed by Eq.~(\ref{LE}), a single-particle wave packet in the strongly dissipative lattice undergoes a rapid decay of the component outside the lower band, after which no subsequent dynamics occurs inside the QZ subspace. Nevertheless, such a decay can cause sudden recoil of the wave packet in both real and momentum spaces \cite{SMA}.

To trigger a QZ dynamics \cite{Facchi2000}, we apply a linear potential $\hat H=-\boldsymbol{F}\cdot\hat{\boldsymbol{r}}$, where $\hat{\boldsymbol{r}}=\sum_{\boldsymbol{r},\sigma}\boldsymbol{r}\hat c^\dag_{\boldsymbol{r}\sigma}\hat c_{\boldsymbol{r}\sigma}$ is the position operator and $\boldsymbol{F}$ is a constant vector. Without dissipation, the COM dynamics 
is described by $\langle\dot{\hat{\boldsymbol{r}}}\rangle=\frac{i}{\hbar}\langle[\hat H,\hat{\boldsymbol{r}}]\rangle$. Clearly, such a potential-only Hamiltonian cannot cause any displacement of the COM, 
since $[\hat H,\hat{\boldsymbol{r}}]=0$. 

In the presence of strong dissipation (\ref{LE}), the accessible Hilbert space is confined to a single band due to the QZ effect. As a result, the projected position operators may no longer commute with each other \cite{Facchi2015}. At the single-particle level, by denoting the projection onto the lower band as $\hat P_-=\sum_{\boldsymbol{k}}|\boldsymbol{k}-\rangle\langle\boldsymbol{k}-|$ and defining $\hat O_-\equiv\hat P_-\hat O\hat P_-$ for $\forall\;\hat O$, we can show that \cite{Nagaosa2010,Niu2010,SMD}
\begin{equation}
[\hat x_-,\hat y_-]=i\hat\Omega\equiv\sum_{\boldsymbol{k}}i\Omega_{xy}(\boldsymbol{k})|\boldsymbol{k}-\rangle\langle\boldsymbol{k}-|,
\label{noncomu}
\end{equation}
where $\Omega_{xy}(\boldsymbol{k})=i(\langle\partial_{k_x}u_{\boldsymbol{k}}|\partial_{k_y}u_{\boldsymbol{k}}\rangle-\langle\partial_{k_y}u_{\boldsymbol{k}}|\partial_{k_x}u_{\boldsymbol{k}}\rangle)$ is the Berry curvature with $|u_{\boldsymbol{k}}\rangle\equiv e^{-i\boldsymbol{k}\cdot\hat{\boldsymbol{r}}}|\boldsymbol{k}-\rangle$ being a Bloch state in the lower band. On the other hand, $[\hat{\boldsymbol{r}}_-, \hat{\boldsymbol{k}}_-]_{\mu\nu}=i\delta_{\mu\nu}$ ($\mu,\nu=x,y$) stays the same as the commutation relation in unconstrained space.

Based on Eq.~(\ref{noncomu}), we can easily write down the projected equations of motion $\langle\dot{\hat{\boldsymbol{k}}}\rangle=\frac{i}{\hbar}\langle[\hat H_-,\hat{\boldsymbol{k}}_-]\rangle=\frac{\boldsymbol{F}}{\hbar}$ and $\langle\dot{\hat{\boldsymbol{r}}}\rangle=\frac{i}{\hbar}\langle[\hat H_-,\hat{\boldsymbol{r}}_-]\rangle=-\frac{\boldsymbol{F}}{\hbar}\times\boldsymbol{e}_z\langle\hat\Omega\rangle$, which 
well approximates the entire open quantum system dynamics in the QZ regime \cite{Zanardi2014,Jiang2016}. If the wave packet is well localized in both real and momentum spaces with the centers of mass being $\boldsymbol{r}$ and $\boldsymbol{k}$, respectively, the semiclassical description of the dynamics follows \cite{Nagaosa2010,Niu2010}:
\begin{equation}
\hbar\dot{\boldsymbol{k}}=\boldsymbol{F},\;\;\;\;
\dot{\boldsymbol{r}}=-\dot{\boldsymbol{k}}\times\boldsymbol{\Omega}(\boldsymbol{k}),
\label{semi}
\end{equation}
where $\boldsymbol{\Omega}(\boldsymbol{k})\equiv\Omega_{xy}(\boldsymbol{k})\boldsymbol{e}_z$. The velocity $\dot{\boldsymbol{r}}$ turns out to be orthogonal to the potential gradient $\boldsymbol{F}$, implying a nonzero Hall conductance. Furthermore, the second equation in Eq.~(\ref{semi}) can be rewritten as $d\boldsymbol{r}=-d\boldsymbol{k}\times\boldsymbol{\Omega}(\boldsymbol{k})$, with the time argument eliminated. 
This is a manifestation of the geometric nature of the QZ dynamics \cite{Zanardi2015,Jiang2016}.

We note that the \emph{bare} lattice system has a trivial band structure, since the intrinsic hopping is suppressed by assumption. The nontrivial wave-packet dynamics (\ref{semi}) is caused by the QZ effect. This is in sharp contrast to the usual anomalous Hall effect, which comes 
solely from the intrinsic Hamiltonian \cite{Nagaosa2010,Niu2010}. While the ZHE (\ref{semi}) is apparently equivalent to the \emph{flat-band} anomalous Hall effect, the physical mechanisms are completely different. The flatness simply inherits from the trivial bare Hamiltonian as an insulator with vanishing 
intrinsic hopping. While it is by no means easy to design an intrinsic flat band \cite{Tasaki2003,Katsura2010,Altman2010}, the QZ subspace naturally serves as a rigorous flat band \cite{Lesanovsky2014}. We will show that a flat band with a nontrivial Berry curvature causes the novel retroreflection at the edge (see Fig.~\ref{fig1} (d) and Fig.~\ref{fig4} (a)).

\emph{Numerical demonstration.---} To directly demonstrate the ZHE, we perform numerical simulations of the Lindblad equation $\dot{\hat\rho}=-\frac{i}{\hbar}[\hat H,\hat\rho]+\Gamma\sum_{\boldsymbol{r}}(\hat L_{\boldsymbol{r}}\hat\rho\hat L^\dag_{\boldsymbol{r}}-\frac{1}{2}\{\hat L^\dag_{\boldsymbol{r}}\hat L_{\boldsymbol{r}},\hat\rho\})$, where the coherent perturbation $\hat H=-\boldsymbol{F}\cdot\hat{\boldsymbol{r}}$ is introduced. At the single-particle level, the equation of motion can be simplified as  
\begin{equation}
i\hbar\partial_t|\tilde\psi\rangle=\hat H_{\rm eff}|\tilde\psi\rangle,
\label{effSchro}
\end{equation}
where $\hat H_{\rm eff}\equiv\hat H-\frac{i}{2}\hbar\Gamma\sum_{\boldsymbol{r}}\hat L^\dag_{\boldsymbol{r}}\hat L_{\boldsymbol{r}}$ is the \emph{non-Hermitian} effective Hamiltonian. Then $\hat\rho$ is given by $\hat\rho=|\tilde\psi\rangle\langle\tilde\psi|+(1-\langle\tilde\psi|\tilde\psi\rangle)|{\rm vac}\rangle\langle{\rm vac}|$ due to 
quantum jumps \cite{Dalibard1992,Plenio1998,Daley2014b}. 

\begin{figure}
\begin{center}
        \includegraphics[width=8cm, clip]{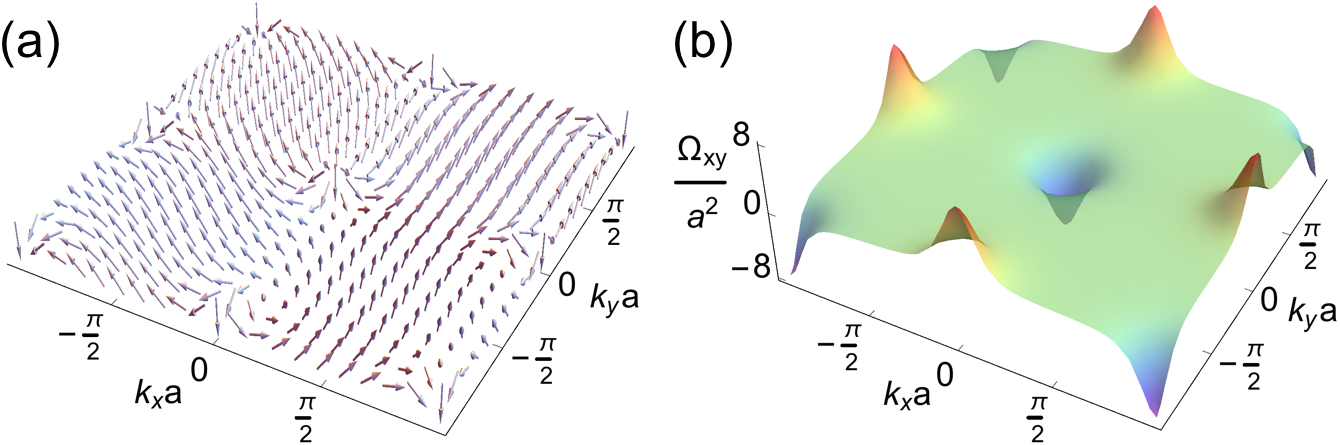}\\
      \end{center}
   \caption{(color online). (a) Spin texture and (b) Berry curvature $\Omega_{xy}(\boldsymbol{k})$ of the lower band (QZ subspace) emerging from the engineered dissipation (\ref{LE}) corresponding to the jump operators in Eq.~(\ref{JO}) with $l_{\uparrow}=-i$.}
   \label{fig2}
\end{figure}

We solve Eq.~(\ref{effSchro}) by exact diagonalization for the lattice size $N=80$. The jump operator is chosen to be 
\begin{equation}
\hat L_{\boldsymbol{r}}=l_{\uparrow}\hat c_{\boldsymbol{r}\uparrow}+\hat c_{\boldsymbol{r}+\boldsymbol{a}_x,\downarrow}+i\hat c_{\boldsymbol{r}+\boldsymbol{a}_y,\downarrow}-\hat c_{\boldsymbol{r}-\boldsymbol{a}_x,\downarrow}-i\hat c_{\boldsymbol{r}-\boldsymbol{a}_y,\downarrow},
\label{JO}
\end{equation}
where the phase difference between the coefficients of $\hat c_{\boldsymbol{r}\pm\boldsymbol{a}_\mu,\downarrow}$ induces a nontrivial Berry curvature. We set $l_{\uparrow}=-i$ in Eq.~(\ref{JO}) to ensure a finite damping gap, and the corresponding Berry curvature $\Omega_{xy}(\boldsymbol{k})$ is plotted in Fig.~\ref{fig2} (b). We see that $\Omega_{xy}(\boldsymbol{k})$ takes on large values only near $\boldsymbol{k}=(0,0),(\pi,0),(0,\pi)$ and $(\pi,\pi)$, where the spin texture is significantly twisted (Fig.~\ref{fig2} (a)). Thus we initially prepare a wave packet centered at $\boldsymbol{k}=(0,0)$ for easy observation of the ZHE, and adjust the wave-packet spread to be $\sigma^2_r=\frac{Na^2}{4\pi}$, so that the size 
relative to the full space stays unchanged after Fourier transformation into momentum space. 

To reach the QZ regime, we choose $\Gamma=10^3\omega$ such that $\Gamma$ is much larger than the Bloch frequency $\omega=\frac{Fa}{\hbar}$, which is the only energy scale of the bare lattice system. Such strong dissipation simplifies the state initialization --- a Gaussian packet prepared in the full space will automatically be projected into the QZ subspace after a very short time $O(\Gamma^{-1})$. The simplicity of the initial state preparation would also be a unique advantage of our scheme for experimental implementations.
 
\begin{figure}
\begin{center}
        \includegraphics[width=8cm, clip]{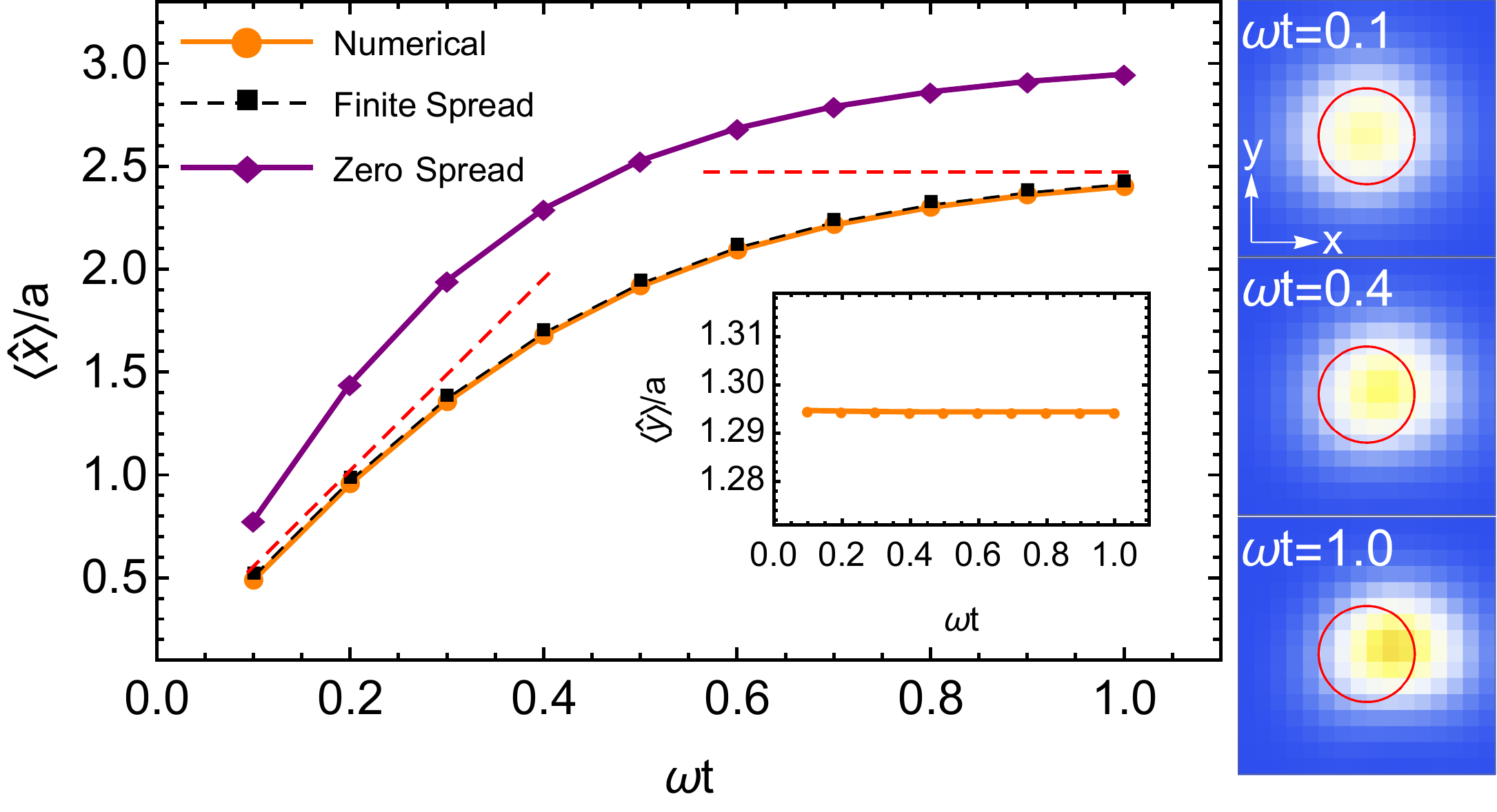}\\
      \end{center}
   \caption{(color online). Time evolution of the $x$-component of the COM
   , obtained from exact diagonalization of Eq.~(\ref{effSchro}) (orange), the integral of the Berry curvature in Eq.~(\ref{integral}) with respect to the trajectory in the Brillouin zone (purple), and the integral of the smoothed Berry curvature that takes into account a finite wave-packet spread (black). The initial increment and the final saturation reflect the Berry curvature landscape (see Fig.~\ref{fig2}). Three panels on the right show the real-space density profiles of the 
   wave packet at $\omega t=0.1,0.4,$ and $1.0$, showing the transverse COM motion, where $\omega=\frac{Fa}{\hbar}$. The inset shows the displacement along the $y$-axis, which stays unchanged and takes on a nonzero value due to the initial recoil \cite{SMA}. The red dashed lines in the main panel and the circles in the right panel show guides to the eye.}
   \label{fig3}
\end{figure}

\begin{figure}
\begin{center}
        \includegraphics[width=8cm, clip]{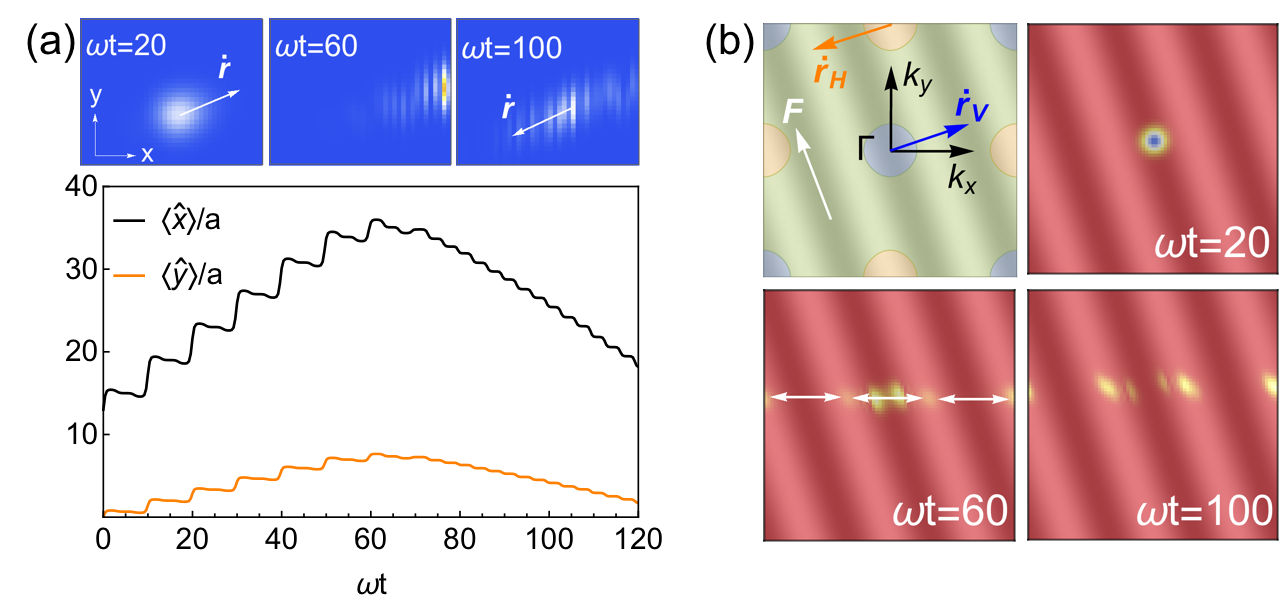}\\
      \end{center}
   \caption{(color online). (a) Long-time dynamics of the COM. 
   Three panels on the top show the real-space density profiles of the wave packet before ($\omega t=20$), at ($\omega t=60$) and after ($\omega t=100$) collision with the right boundary, indicating the retroreflection. (b) ``Hills" (orange, $\Omega_{xy}>1$) and ``valleys'' (blue, $\Omega_{xy}<-1$) in the Brillouin zone and the wave-packet profile in momentum space before, at and after the retroreflection. Dark (Light) stripes, which are parallel to $\boldsymbol{F}$, refer to the orbits in which the anomalous motion in real space approaches (leaves) the right boundary.}
   \label{fig4}
\end{figure}

\emph{Short-time dynamics.---} For a short-time interval $[0,\omega^{-1}]$, $\langle\hat\Omega\rangle$ deviates significantly from zero. For $\boldsymbol{F}=F\boldsymbol{e}_y$, the time evolution of the center of mass (COM) along the $x$-axis is plotted in Fig.~\ref{fig3}, where stroboscopic snapshots of the wave packet are also presented, showing transverse displacements. Qualitatively, the slowdown behavoir is consistent with a decrease in the Berry curvature (Fig.~\ref{fig2}) when $k_x=0$ and $k_y$ increases from $0$. Quantitatively, however, the numerical result (orange curve in Fig.~\ref{fig3}) clearly deviates from the semiclassical theory (\ref{semi}) (purple curve in Fig.~\ref{fig3}), which is evaluated by
\begin{equation}
\langle\hat x\rangle=-\int^{k_y}_0 dk'_y\Omega_{xy}(k'_x=0,k'_y),
\label{integral}
\end{equation}
where $k_ya=\omega t$. Such a deviation arises mainly from a finite spread $\sigma^2_k=\frac{\pi}{Na^2}$ of the wave packet in momentum space, which can be shown to stay (approximately) Gaussian \cite{SMA}. We take into account a finite spread by smoothing the Berry curvature via convolution with a Gaussian kernel: $\tilde\Omega_{xy}(\boldsymbol{k})=\frac{1}{2\pi\sigma^2_k}\int_{\rm B.Z.} d^2\boldsymbol{k}'e^{-(\boldsymbol{k}'-\boldsymbol{k})^2/(2\sigma^2_k)}\Omega_{xy}(\boldsymbol{k}')$. Then, we compute Eq.~(\ref{integral}) in terms of $\tilde\Omega_{xy}(\boldsymbol{k})$ rather than $\Omega_{xy}(\boldsymbol{k})$, and find excellent agreement (see the black curve in Fig.~\ref{fig3}). While $\langle\hat y\rangle$ stays unchanged as expected (inset in Fig.~\ref{fig3}), it takes on a nonzero value due to a recoil resulting from the initial rapid decay of the wave packet outside the lower band \cite{SMA}.

\emph{Long-time dynamics.---} A wave packet with nonzero anomalous velocity can reach a boundary. For example, when $\boldsymbol{F}=\frac{F}{\sqrt{10}}(-1,3)$ with rational ratio $\frac{F_y}{F_x}=-3$, the trajectory in the Brillouin zone forms a closed orbit which mainly crosses the ``valleys'' in Fig.~\ref{fig2} (b) with negative $\Omega_{xy}$. Nevertheless, since $\Omega_{xy}$ is small in a large area of the Brillouin zone, we expect that the COM 
moves in a stepwise manner, as confirmed numerically in Fig.~\ref{fig4} (a). Surprisingly, after collision with the right boundary, the wave packet is \emph{retroreflected} \cite{MV}. This is a unique feature of a flat band with a nontrivial Berry curvature, where the real-space dynamics is dominated by the anomalous velocity, which is always perpendicular to $\boldsymbol{F}$ in Eq.~(\ref{semi}) but with only two possible directions, depending on the sign of $\Omega_{xy}$. As shown in Fig.~\ref{fig4} (b), the wave packet in the Brillouin zone diffuses from the orbits crossing the ``valleys'' (dark stripes) to those crossing the ``hills'' (light stripes) during the collision, due to the compression in real space (see the middle panel in Fig.~\ref{fig4} (a)), leading to inversion of the anomalous velocity. After the retroreflection, the wave packet fragments into three pieces in the Brillouin zone, shortening the step length of motion to one-third of that before retroreflection (see the $\langle\hat x\rangle,\langle\hat y\rangle - t$ curves in Fig.~\ref{fig4} (a)).

We note that the Chern number of a perfectly flat band always vanishes if $\hat L_{\boldsymbol{r}}$'s are short-ranged \cite{Tang2014,Diehl2015}, implying the existence of ``hills'' whenever there are ``valleys''. Furthermore, we can argue that the wave packet always diffuses from ``valleys" to ``hills" (or vice versa) to reverse the anomalous velocity by colliding with a boundary \cite{SMM}. We thus expect the retroreflection to be universal, although the wave-packet profile after the retroreflection is model-dependent.
 
In the edge physics of the usual anomalous Hall effect, a particle collides with a boundary at a \emph{normal group velocity}, so \emph{reflection} occurs accompanied by lateral displacement, known as a matter-wave analogy of the Goos-H\"anchen's shift \cite{Niu2005b}. For the quantum anomalous Hall effect \cite{Haldane1988}, a state highly localized at the edge can undergo persistent lateral motion with definite chirality, in analogy with the skipping motion of a charged particle in the presence of magnetization \cite{Qi2016}. In contrast, collision with a boundary by an \emph{anomalous velocity alone} leads to retroreflection.

\emph{Physical implementation.---} With excellent scalability and controllability \cite{Bloch2008}, ultracold atomic systems are promising candidates for implementing the ZHE. A crucial ingredient is a dissipative optical lattice with collective one-body loss, which can be realized by a combination of a fine-tuned Rabi coupling and on-site one-body loss of an auxiliary internal state. The real-space dynamics can be directly visualized by means of quantum-gas microscopy \cite{Greiner2009}. The retroreflection can also be tested by using a box potential \cite{Hadzibabic2013}.

\begin{figure}
\begin{center}
        \includegraphics[width=8cm, clip]{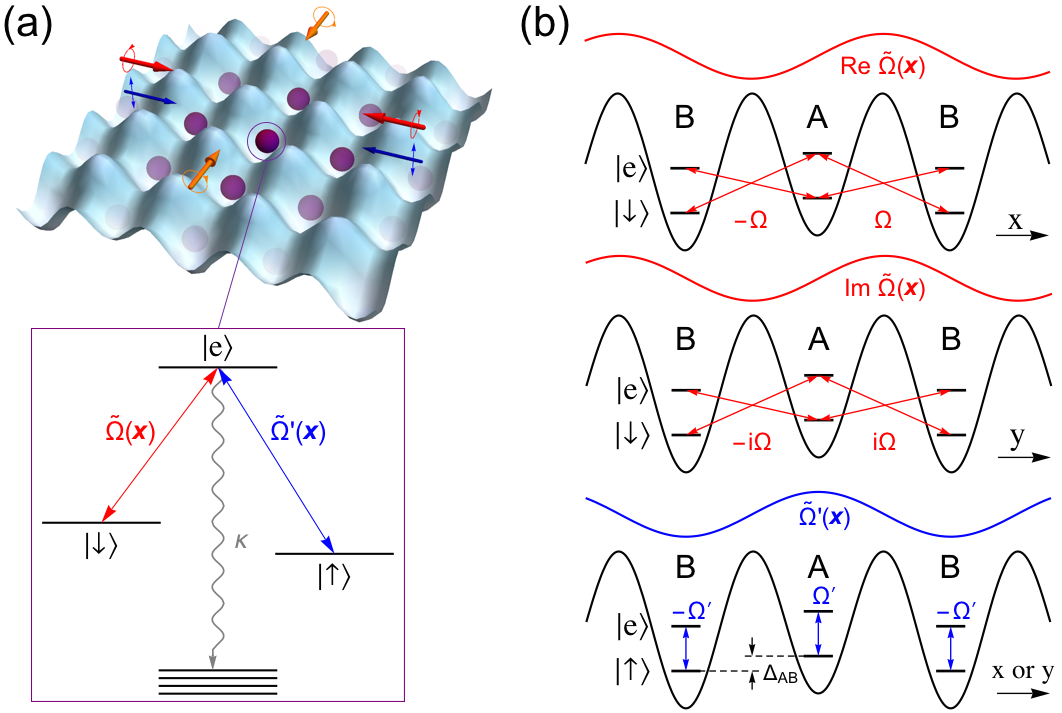}\\
      \end{center}
   \caption{(color online). (a) $\Lambda$ atoms in a staggered square optical lattice subjected to a set of fine-tuned lasers, which couple internal states $|\uparrow\rangle$ and $|\downarrow\rangle$ to $|e\rangle$ with  spatially dependent Rabi frequencies $\tilde\Omega(\boldsymbol{x})$ and $\tilde\Omega'(\boldsymbol{x})$, respectively. The excited state $|e\rangle$ is unstable and undergoes effectively a rapid on-site loss at a rate $\kappa$. (b) Within the tight-binding approximation, $|e\rangle$ is coupled to the nearest $|\downarrow\rangle$'s in a $p$-wave symmetry and the on-site $|\uparrow\rangle$ with a phase inversion for the sublattices $A$ and $B$. The magnitudes of the Rabi frequencies for $|\uparrow\rangle\leftrightarrow|e\rangle$ and $|\downarrow\rangle\leftrightarrow|e\rangle$ transitions are $|\Omega|$ and $|\Omega'|$, respectively. The energy levels are not to scale.}
   \label{fig5}
\end{figure}

Let us discuss how to engineer the dissipation (\ref{JO}) used for numerical demonstration. We consider $\Lambda$ atoms with two ground states denoted by $|\uparrow\rangle$ and $|\downarrow\rangle$ and an excited state denoted by $|e\rangle$ in a staggered square optical lattice (Fig.~\ref{fig5} (a)) \cite{Bloch2015}. The on-site energy difference between the sublattice $A$ and $B$, denoted by $\Delta_{AB}$, is set to be much larger than the nearest-neighbor tunneling, so as to suppress the kinetic degree of freedom \cite{Dalibard2011,Ketterle2013,Bloch2013,Goldman2014,Bloch2015}. While $|\uparrow\rangle$ and $|\downarrow\rangle$ are stable, $|e\rangle$ quickly decays into external states at an effective on-site loss rate $\kappa$. By applying a set of fine-tuned lasers \cite{SME}, $|e\rangle$ is coupled to the nearest $|\downarrow\rangle$ in a $p$-wave symmetry, i.e., with the Rabi frequencies being $\Omega$, $i\Omega$, $-\Omega$, $-i\Omega$ in the counterclockwise direction, and also coupled to the on-site $|\uparrow\rangle$ with a Rabi frequency $\pm\Omega'$ in a checkerboard pattern (Fig.~\ref{fig5} (b)). If $\kappa\gg|\Omega|,|\Omega'|$, we can adiabatically eliminate the fast decay mode $|e\rangle$ \cite{Sorensen2012} to obtain a closed equation of motion (\ref{LE}) that involves only the internal states $|\uparrow\rangle$ and $|\downarrow\rangle$ with $\Gamma=\frac{|\Omega|^2}{\kappa}$ in Eq.~(\ref{LE}) and $l_\uparrow=\frac{\Omega'}{\Omega}$ in Eq.~(\ref{JO}) \cite{SME}. A detailed setup based on $^{87}{\rm Rb}$ is available in the Supplemental Material.

\emph{Summary and outlook.---} We have predicted a unique Hall effect due to dissipation --- the ZHE based on a general scheme to tailor the Hilbert space into a Bloch band by using the QZ effect. Whenever the band is flat and possesses a nonzero Berry curvature, the QZ dynamics of a wave packet exhibits a transverse velocity in response to a potential gradient and the \emph{retroreflection} on a boundary. We have also explored a possible implementation of the ZHE with ultracold atoms.

Our work opens up a possibility of investigating many-body physics inside a QZ subspace, which can naturally be made flat \cite{Tasaki2003,Katsura2010,Altman2010}. With tunable interaction \cite{Chin2010}, ultracold atoms provide an ideal platform for experimental implementation. Even for free particles, we expect nontrivial generalizations to other lattice structures, Bogoliubov quasiparticles \cite{Diehl2012,Molmer2015,Molmer2016}, spatiotemporal potential gradient \cite{Niu1998}, multiple \cite{Shindou2005} and topologically nontrivial bands \cite{Goldman2016}, and noiseless subsystems \cite{Viola2000}. A systematic development of a QZ effect-based toolbox for manipulating wave packets will offer a unique possibility for quantum information processing with continuous variables \cite{Braunstein2005}. 

We acknowledge Y. Ashida, S. Choi, I. Danshita, T. Fukuhara, S. Furukawa, R. Hamazaki, H. Katsura and J. Schmiedmayer for valuable discussions. This work was supported by KAKENHI Grant No. 26287088 from the Japan Society for the Promotion of Science, a Grant-in-Aid for Scientific Research on Innovative Areas ``Topological Materials Science" (KAKENHI Grant No. 15H05855), and the Photon Frontier Network Program from MEXT of Japan. Z. G. was supported by MEXT. S. H. acknowledges support from JSPS (Grant No. 16J03619) and through the Advanced Leading Graduate Course for Photon Science (ALPS).

\bibliography{GZP_references}

\clearpage
\begin{center}
\textbf{\large Supplemental Materials}
\end{center}
\setcounter{equation}{0}
\setcounter{figure}{0}
\setcounter{table}{0}
\makeatletter
\renewcommand{\theequation}{S\arabic{equation}}
\renewcommand{\thefigure}{S\arabic{figure}}
\renewcommand{\bibnumfmt}[1]{[S#1]}

Here we provide the derivations Eqs.~(2) and (3) in the main text, additional numerical results on the short-time dynamics, further information on the retroreflection and the details of the experimental implementation discussed in the main text.

\section{Derivation of Eq.~(2)}
A Lindblad master equation is invariant under an arbitrary unitary transformation of the jump operators \cite{Breuer2002}, that is
\begin{equation}
\sum_j\mathcal{D}[\hat L_j]\hat\rho=\sum_j\mathcal{D}[\hat L'_j]\hat\rho,
\label{lindu}
\end{equation}
where $\mathcal{D}[\hat L]\hat\rho\equiv\hat L\hat\rho\hat L^\dag-\{\hat L^\dag\hat L,\hat\rho\}/2$ and $\hat L'_j=\sum_k U_{jk}\hat L'_k$, with $U_{jk}$ being an arbitrary $c$-number unitary matrix. Equation (\ref{lindu}) 
can straightforwardly be checked by using the unitarity of $U_{jk}$, i.e., $\sum_k U_{jk}U^*_{lk}=\delta_{jl}$.

In particular, when the subscript $j$ denotes a site $\boldsymbol{r}$ of an $N\times N$ lattice, we can choose the unitary transformation to be the Fourier transformation, i.e., $U_{\boldsymbol{k}\boldsymbol{r}}=N^{-1}e^{-i\boldsymbol{k}\cdot\boldsymbol{r}}$, where $\boldsymbol{k}$ is the crystal momentum. We write the translationally invariant jump operator 
as $\hat L_{\boldsymbol{r}}=\sum_{\boldsymbol{r}'}l_{\boldsymbol{r}'-\boldsymbol{r},\sigma}\hat c_{\boldsymbol{r}'\sigma}$, and obtain
\begin{equation}
\begin{split}
\hat L_{\boldsymbol{k}}&=N^{-1}\sum_{\boldsymbol{r}}e^{-i\boldsymbol{k}\cdot\boldsymbol{r}}\hat L_{\boldsymbol{r}}\\
&=N^{-1}\sum_{\boldsymbol{r},\boldsymbol{r}',\sigma}e^{-i\boldsymbol{k}\cdot\boldsymbol{r}}l_{\boldsymbol{r}'-\boldsymbol{r},\sigma}\hat c_{\boldsymbol{r}'\sigma}\\
&=\sum_{\boldsymbol{r}'-\boldsymbol{r},\sigma}e^{i\boldsymbol{k}\cdot(\boldsymbol{r}'-\boldsymbol{r})}l_{\boldsymbol{r}'-\boldsymbol{r},\sigma}\left(N^{-1}\sum_{\boldsymbol{r}'}e^{-i\boldsymbol{k}\cdot\boldsymbol{r}'}\hat c_{\boldsymbol{r}'\sigma}\right)\\
&=\sum_{\sigma}l_{\boldsymbol{k}\sigma}\hat c_{\boldsymbol{k}\sigma},
\end{split}
\end{equation}
where $l_{\boldsymbol{k}\sigma}\equiv\sum_{\boldsymbol{r}}e^{i\boldsymbol{k}\cdot\boldsymbol{r}}l_{\boldsymbol{r}\sigma}$ and $\hat c_{\boldsymbol{k}\sigma}\equiv N^{-1}\sum_{\boldsymbol{r}}e^{-i\boldsymbol{k}\cdot\boldsymbol{r}}\hat c_{\boldsymbol{r}\sigma}$. 

If we introduce $\Delta_{\boldsymbol{k}}=\sum_\sigma |l_{\boldsymbol{k}\sigma}|^2$ and $\hat c_{\boldsymbol{k}+}=\Delta^{-1/2}_{\boldsymbol{k}}\sum_{\sigma}l_{\boldsymbol{k}\sigma}\hat c_{\boldsymbol{k}\sigma}$, $\hat L_{\boldsymbol{k}}$ can be rewritten as $\Delta^{1/2}_{\boldsymbol{k}}\hat c_{\boldsymbol{k}+}$. By using the fact that $\mathcal{D}[\alpha\hat L]\hat\rho=|\alpha|^2\mathcal{D}[\hat L]\hat\rho$, $\forall\alpha\in\mathbb{C}$, we obtain Eq.~(2) in the main text, i.e.,
\begin{equation}
\dot{\hat\rho}=\Gamma\sum_{\boldsymbol{r}}\mathcal{D}[\hat L_{\boldsymbol{r}}]\hat\rho=\Gamma\sum_{\boldsymbol{k}}\Delta_{\boldsymbol{k}}\mathcal{D}[\hat c_{\boldsymbol{k}+}]\hat\rho.
\end{equation}
Let us define the annihilation operator for the lower band $\hat c_{\boldsymbol{k}-}\equiv\Delta^{-\frac{1}{2}}_{\boldsymbol{k}}(l^*_{\boldsymbol{k}\downarrow}\hat c_{\boldsymbol{k}\uparrow}-l^*_{\boldsymbol{k}\uparrow}\hat c_{\boldsymbol{k}\downarrow})$ such that $\{\hat c^\dag_{\boldsymbol{k}'-},\hat c_{\boldsymbol{k}+}\}=0$ for fermions and $[\hat c^\dag_{\boldsymbol{k}'-},\hat c_{\boldsymbol{k}+}]=0$ for bosons for $\forall\;\boldsymbol{k},\boldsymbol{k}'$. Then all the many-body states, which take the form of $|\Psi_{\{n_{\boldsymbol{k}}\}}\rangle\equiv\prod_{\boldsymbol{k}} (\hat c^\dag_{\boldsymbol{k}-})^{n_{\boldsymbol{k}}}|{\rm vac}\rangle$ with $n_{\boldsymbol{k}}=0,1$ for fermions and $n_{\boldsymbol{k}}\in\mathbb{N}$ for bosons, 
span a decoherence-free subspace \cite{Lidar1998}. In other words, each state with only the emergent lower Bloch band being occupied is decoherence-free, since the necessary and sufficient condition \cite{Lidar1998} $\hat c_{\boldsymbol{k}+}|\Psi_{\{n_{\boldsymbol{k}}\}}\rangle=0$ for $\forall
\;\boldsymbol{k}$ and $\{n_{\boldsymbol{k}}\}$ is satisfied.

\section{Derivation of Eq.~(3)}
Let us calculate the matrix elements of the position operators $\hat x$ and $\hat y$ within the QZ subspace $\mathcal{S}={\rm span}\{|\boldsymbol{k}-\rangle=e^{i\boldsymbol{k}\cdot\hat{\boldsymbol{r}}}|u_{\boldsymbol{k}}\rangle;\boldsymbol{k}\in{\rm B.Z.}\}$. We obtain
\begin{equation}
\begin{split}
&\;\;\;\;\;\langle u_{\boldsymbol{k}}|e^{-i\boldsymbol{k}\cdot\hat{\boldsymbol{r}}}\hat x e^{i\boldsymbol{k}'\cdot\hat{\boldsymbol{r}}}|u_{\boldsymbol{k}'}\rangle\\
&=\langle u_{\boldsymbol{k}}|i\partial_{k_x}e^{i(\boldsymbol{k}'-\boldsymbol{k})\cdot\hat{\boldsymbol{r}}}|u_{\boldsymbol{k}'}\rangle\\
&=i\partial_{k_x}\langle u_{\boldsymbol{k}}|e^{i(\boldsymbol{k}'-\boldsymbol{k})\cdot\hat{\boldsymbol{r}}}|u_{\boldsymbol{k}'}\rangle-i\langle \partial_{k_x}u_{\boldsymbol{k}}|e^{i(\boldsymbol{k}'-\boldsymbol{k})\cdot\hat{\boldsymbol{r}}}|u_{\boldsymbol{k}'}\rangle\\
&=(i\partial_{k_x}+\mathcal{A}_x(\boldsymbol{k}))\delta_{\boldsymbol{k},\boldsymbol{k}'},
\end{split}
\label{nonc}
\end{equation}
where $\mathcal{A}_x(\boldsymbol{k})$ is the $x$-component of the Berry connection $\boldsymbol{\mathcal{A}}(\boldsymbol{k})\equiv i\langle u_{\boldsymbol{k}}|\nabla_{\boldsymbol{k}}u_{\boldsymbol{k}}\rangle$, and the Kronecker delta $\delta_{\boldsymbol{k},\boldsymbol{k}'}$ should be interpreted as $(\frac{2\pi}{Na})^2\delta(\boldsymbol{k}-\boldsymbol{k}')$ in the continuous limit. 
Equation (\ref{nonc}) implies
\begin{equation} 
\hat x_-\sim i\partial_{k_x}+\mathcal{A}_x(\boldsymbol{k})
\end{equation}
in the representation of $|\boldsymbol{k}-\rangle$'s. Similarly, we have
\begin{equation} 
\hat y_-\sim i\partial_{k_y}+\mathcal{A}_y(\boldsymbol{k}).
\end{equation}
Thus the commutator between the projected position operators $\hat x_-$ and $\hat y_-$ should be
\begin{equation}
\begin{split}
[\hat x_-,\hat y_-]&\sim[i\partial_{k_x}+\mathcal{A}_x(\boldsymbol{k}),i\partial_{k_y}+\mathcal{A}_y(\boldsymbol{k})]\\
&=i(\partial_{k_x}\mathcal{A}_y(\boldsymbol{k})-\partial_{k_y}\mathcal{A}_x(\boldsymbol{k}))\equiv i\Omega_{xy}(\boldsymbol{k}),
\end{split}
\label{xy}
\end{equation}
where $\Omega_{xy}(\boldsymbol{k})$ is the Berry curvature. Equation (2) in the main text is nothing but the operator form of Eq.~(\ref{xy}).

\section{Wave-packet recoil}
Let us quantitatively discuss the recoil phenomenon that can be seen in the inset of Fig.~3 in the main text. 
We first expand the initial state vector, which is the direct product of the Gaussian packet in real space and the spin state $\frac{1}{\sqrt{2}}(|\uparrow\rangle+|\downarrow\rangle)$, in terms of $|\boldsymbol{k}\sigma\rangle$'s:
\begin{equation}
\begin{split}
&|\psi\rangle=\int\frac{d^2\boldsymbol{k}}{S_k}e^{-\frac{\boldsymbol{k}^2}{4\sigma^2_{k}}}\frac{\sqrt{\pi}}{Na\sigma_k}(|\boldsymbol{k}\uparrow\rangle+|\boldsymbol{k}\downarrow\rangle)\\
&=\int\frac{d^2\boldsymbol{k}}{S_k}e^{-\frac{\boldsymbol{k}^2}{4\sigma^2_{k}}}\frac{(l_{\boldsymbol{k}\uparrow}+l_{\boldsymbol{k}\downarrow})|\boldsymbol{k}+\rangle+(l^*_{\boldsymbol{k}\downarrow}-l^*_{\boldsymbol{k}\uparrow})|\boldsymbol{k}-\rangle}{Na\sigma_k\sqrt{\Delta_{\boldsymbol{k}}/\pi}},
\end{split}
\end{equation}
where $S_k=(\frac{2\pi}{Na})^2$ and $\sigma^2_k=\frac{\pi}{Na^2}$. This Gaussian wave packet is centered at the origin in either real or momentum space. 
After a rapid decay of the $|\boldsymbol{k}+\rangle$ components, the state vector becomes
\begin{equation}
|\tilde\psi\rangle=\int\frac{d^2\boldsymbol{k}}{S_k}\sqrt{\frac{\pi}{\Delta_{\boldsymbol{k}}}}e^{-\frac{\boldsymbol{k}^2}{4\sigma^2_{k}}}\frac{l^*_{\boldsymbol{k}\downarrow}-l^*_{\boldsymbol{k}\uparrow}}{Na\sigma_k}|\boldsymbol{k}-\rangle,
\label{afterdecay}
\end{equation}
where the tilde indicates that the state is not normalized. If $l_{\boldsymbol{k}\uparrow}$ and $l_{\boldsymbol{k}\downarrow}$ possess different parity symmetries (e.g., $s$-wave and $p$-wave), we have $\langle\tilde\psi|\tilde\psi\rangle=\frac{1}{2}$, and the expectation value of the crystal momentum is evaluated to be
\begin{equation}
\langle\hat{\boldsymbol{k}}\rangle=\frac{\langle\tilde\psi|\hat{\boldsymbol{k}}|\tilde\psi\rangle}{\langle\tilde\psi|\tilde\psi\rangle}=-\int\frac{d^2\boldsymbol{k}}{\pi\sigma^2_k\Delta_{\boldsymbol{k}}}{\rm Re}(l_{\boldsymbol{k}\uparrow}l^*_{\boldsymbol{k}\downarrow})e^{-\frac{\boldsymbol{k}^2}{2\sigma^2_{k}}}\boldsymbol{k},
\label{expk}
\end{equation}
which is invariant under the gauge transformation $l_{\boldsymbol{k}\sigma}\to e^{i\alpha(\boldsymbol{k})}l_{\boldsymbol{k}\sigma}$, where $\alpha(\boldsymbol{k})$ is an arbitrary real function of $\boldsymbol{k}$. For the choice of parameters in the main text, we have $l_{\boldsymbol{k}\uparrow}=l_\uparrow$, $l_{\boldsymbol{k}\downarrow}=2i\sin k_xa-2\sin k_ya$ and $\Delta_{\boldsymbol{k}}=|l_\uparrow|^2+4(\sin^2k_xa+\sin^2k_ya)$; thus Eq.~(\ref{expk}) gives $\langle\hat{\boldsymbol{k}}\rangle a=\mp0.102\boldsymbol{e}_x$ for $l_\uparrow=\pm i$ and $\langle\hat{\boldsymbol{k}}\rangle a=\pm0.102\boldsymbol{e}_y$ for $l_\uparrow=\pm1$. While such a small recoil in momentum space has negligible influence on the wave-packet dynamics for $l_\uparrow=\pm i$, it leads to an observable consequence for $l_\uparrow=\pm1$, as we will see from the numerical results (Fig.~\ref{figS3}).


To calculate the displacement in real space, we further rewrite Eq.~(\ref{afterdecay}) as
\begin{equation}
|\tilde\psi\rangle=\int\frac{d^2\boldsymbol{k}}{S_k}\frac{\sqrt{\pi}}{\Delta_{\boldsymbol{k}}}e^{-\frac{\boldsymbol{k}^2}{4\sigma^2_{k}}}\frac{l^*_{\boldsymbol{k}\downarrow}-l^*_{\boldsymbol{k}\uparrow}}{Na\sigma_k}(l_{\boldsymbol{k}\downarrow} |\boldsymbol{k}\uparrow\rangle-l_{\boldsymbol{k}\uparrow} |\boldsymbol{k}\downarrow\rangle).
\end{equation}
By defining $\tilde\psi_\sigma(\boldsymbol{k})\equiv \frac{Na}{2\pi}\langle\boldsymbol{k}\sigma|\tilde\psi\rangle$ and the spin-flip notation $\bar\sigma$ (e.g., $\bar\sigma=\downarrow$ if $\sigma=\uparrow$), we obtain 
\begin{equation}
\tilde\psi_\sigma(\boldsymbol{k})=\frac{1}{2\sqrt{\pi}\sigma_k}\frac{|l_{\boldsymbol{k}\bar\sigma}|^2-l_{\boldsymbol{k}\bar\sigma}l^*_{\boldsymbol{k}\sigma}}{\Delta_{\boldsymbol{k}}}e^{-\frac{\boldsymbol{k}^2}{4\sigma^2_{k}}}.
\end{equation}
We can use this result to evaluate the center of mass $\langle\hat{\boldsymbol{r}}\rangle=2\sum_\sigma\int d^2\boldsymbol{k}\tilde\psi^*_\sigma(\boldsymbol{k})i\nabla_{\boldsymbol{k}}\tilde\psi_\sigma(\boldsymbol{k})$ as
\begin{equation}
\begin{split}
\langle\hat{\boldsymbol{r}}\rangle=\sum_\sigma\int\frac{d^2\boldsymbol{k}}{2\pi\sigma^2_k}\frac{|l_{\boldsymbol{k}\sigma}|^2-l^*_{\boldsymbol{k}\sigma}l_{\boldsymbol{k}\bar\sigma}}{\Delta_{\boldsymbol{k}}}e^{-\frac{\boldsymbol{k}^2}{4\sigma^2_{k}}}\\
\times i\nabla_{\boldsymbol{k}}\left(\frac{|l_{\boldsymbol{k}\sigma}|^2-l_{\boldsymbol{k}\sigma}l^*_{\boldsymbol{k}\bar\sigma}}{\Delta_{\boldsymbol{k}}}e^{-\frac{\boldsymbol{k}^2}{4\sigma^2_{k}}}\right).
\end{split}
\label{avgr}
\end{equation}
Equation (\ref{avgr}) can be simplified as 
\begin{equation}
\begin{split}
\langle\hat{\boldsymbol{r}}\rangle=-\int\frac{d^2\boldsymbol{k}}{\pi\sigma^2_k\Delta_{\boldsymbol{k}}}{\rm Im}(l_{\boldsymbol{k}\uparrow}l^*_{\boldsymbol{k}\downarrow})e^{-\frac{\boldsymbol{k}^2}{4\sigma^2_{k}}}\\
\times\nabla_{\boldsymbol{k}}\left(\frac{|l_{\boldsymbol{k}\uparrow}|^2-|l_{\boldsymbol{k}\downarrow}|^2}{\Delta_{\boldsymbol{k}}}e^{-\frac{\boldsymbol{k}^2}{4\sigma^2_{k}}}\right),
\end{split}
\label{avgrs}
\end{equation}
which is again gauge invariant. For the specific choice of parameters in the main text, Eq.~(\ref{avgrs}) gives 
$\langle\hat{\boldsymbol{r}}\rangle=\pm1.29495a\boldsymbol{e}_y$ for $l_\uparrow=\mp i$ and $\langle\hat{\boldsymbol{r}}\rangle=\pm1.29495a\boldsymbol{e}_x$ for $l_\uparrow=\mp1$, which are quantitatively consistent with the numerical result $|\langle\hat{\boldsymbol{r}}\rangle|=1.29475$ obtained from exact diagonalization.


\section{Additional numerical results on the short-time dynamics}
Here we present extensive numerical results to get more physical and intuitive insights into the ZHE. From now on, unless specified otherwise, the parameters  are the same as those used for the numerical demonstration in the main text.
 
\subsection{Influence of the coupling strength $\Gamma$}
\begin{figure}
\begin{center}
        \includegraphics[width=7cm, clip]{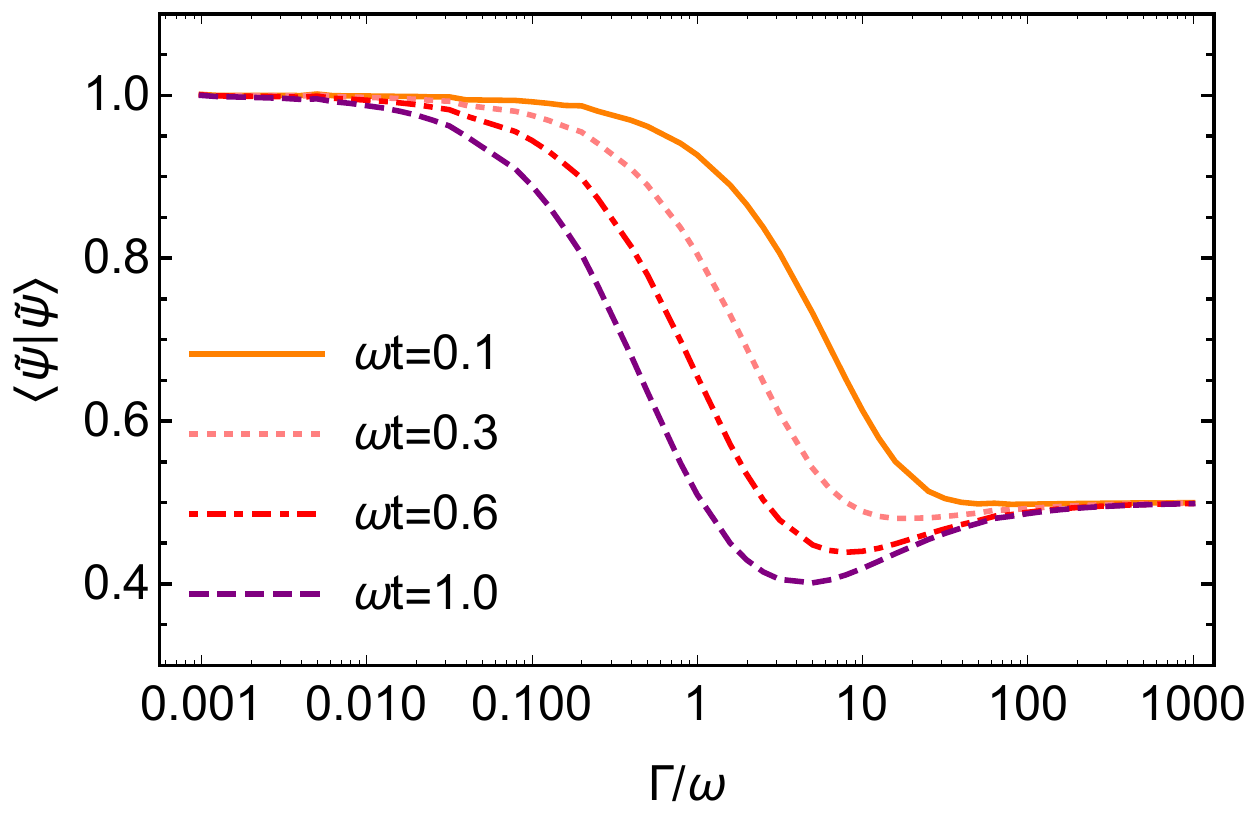}\\
      \end{center}
   \caption{Loss behavior quantified by the dependence of $\langle\tilde\psi|\tilde\psi\rangle$ on $\Gamma$. While $\langle\tilde\psi|\tilde\psi\rangle$ monotonically decreases as $t$ increases, it is not monotonic with respect to $\Gamma$ due to the QZ effect. The lattice size is chosen to be $N=40$.}
   \label{figS1}
\end{figure}

While it is clear from the discussions in the main text that $\Gamma=10^3\omega$ leads to almost the same results in the $\Gamma\to\infty$ limit, the magnitude of $\Gamma$  leading to the onset of the ZHE remains to be determined. To estimate it, we make use of the $\langle\tilde\psi|\tilde\psi\rangle$ -- $\Gamma$ relation plotted in Fig.~\ref{figS1}, since the suppression of loss is a hallmark of the QZ effect \cite{Cirac2008,Ott2013,Rey2014}. At time $t=\omega^{-1}$ (purple curve), the minimum of $\langle\tilde\psi|\tilde\psi\rangle$ is found to lie in the range $1<\Gamma/\omega<10$; thus we expect that the ZHE is evident even for $\Gamma\sim10\omega$. 

\begin{figure}
\begin{center}
        \includegraphics[width=8cm, clip]{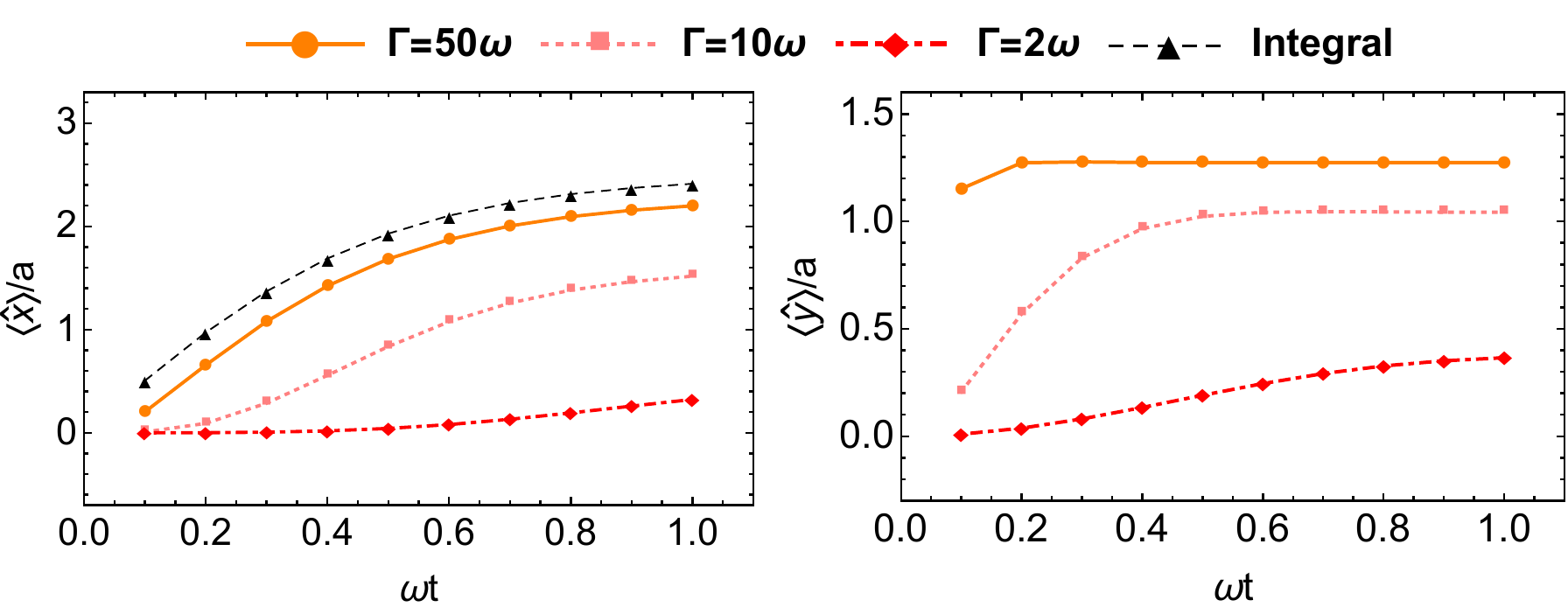}\\
      \end{center}
   \caption{Time evolution of the center of mass for $\Gamma=2\omega$ (red dash-dotted), $10\omega$ (pink dotted) and $50\omega$ (orange solid). Here $\langle\hat{x}\rangle$ shows the transverse displacement. The values in the quantum Zeno limit ($\Gamma\to\infty$), obtained by integrating the smoothed Berry curvature (black dashed), are also presented.}
   \label{figS2}
\end{figure}

Such an expectation is confirmed by the direct calculation for the time evolutions of the center of mass, which are shown in Fig.~\ref{figS2}. It is found that at $t=\omega^{-1}$, $\langle\hat x\rangle$ reaches $1.5a$ for $\Gamma=10\omega$ (pink curve) and $2.2a$ for $\Gamma=50\omega$ (orange curve), and that the latter is close to the limiting value $2.4a$ (black curve). On the other hand, we can see little displacement for $\Gamma=2\omega$ (red curve), where the dissipation is so weak that the QZ effect is not significant. 

\subsection{Influence of the wave-packet spread}
As mentioned in the main text, the discrepancy between the theoretical prediction in the semiclassical limit and the numerical exact-diagonalization result arises from a finite spread of the wave packet in the Brillouin zone (BZ). After smoothing the Berry curvature and redoing the calculation, we have obtained agreement between the two results (see Fig. 3 in the main text). On the other hand, we expect that the semiclassical prediction can be reproduced by reducing the spread in the BZ, or equivalently by making the wave packet wider in real space. 

\begin{figure}
\begin{center}
        \includegraphics[width=7cm, clip]{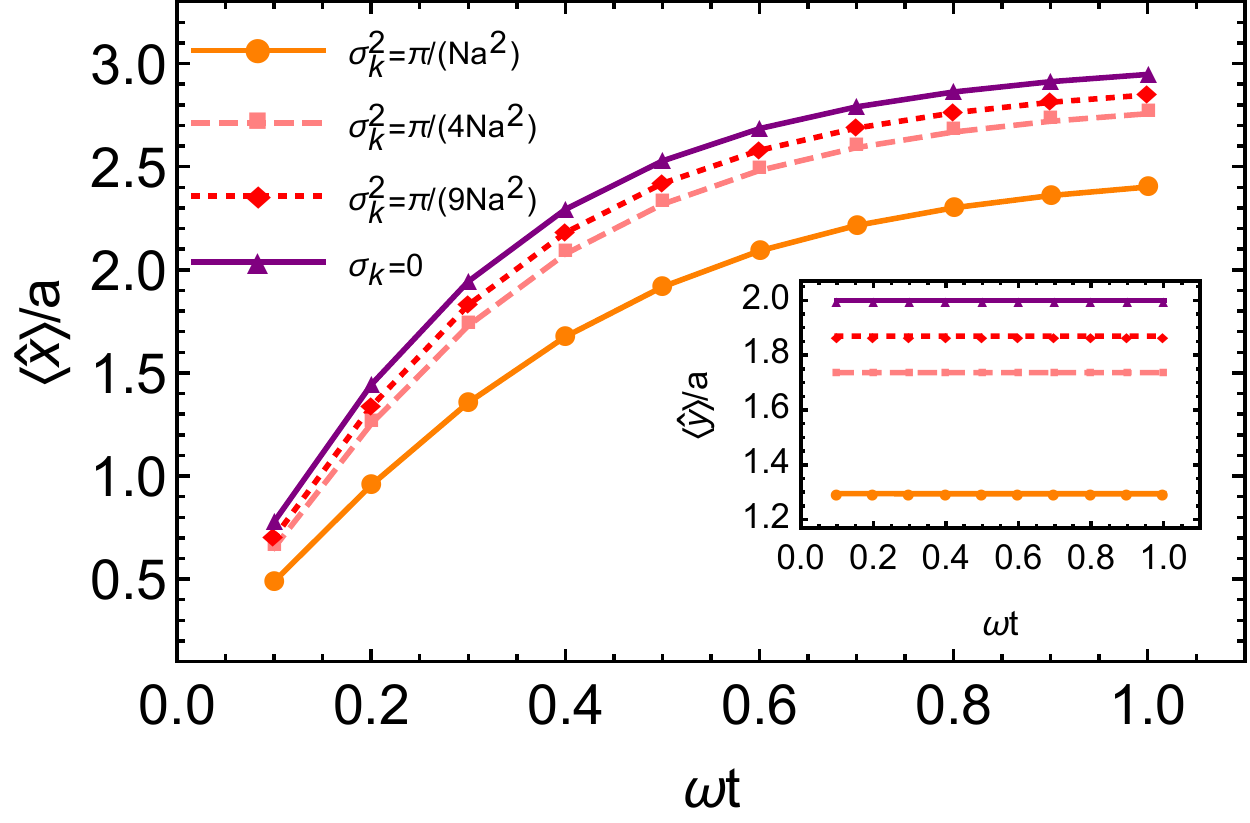}\\
      \end{center}
   \caption{Time evolution of the center of mass for different momentum-space spreads $\sigma_k=\sqrt{\frac{\pi}{N}}\frac{1}{a}$ (orange), $\sqrt{\frac{\pi}{N}}\frac{1}{2a}$ (pink) and $\sqrt{\frac{\pi}{N}}\frac{1}{3a}$ (red) corresponding to the real-space spreads $\sigma_r=\sqrt{\frac{N}{\pi}}\frac{a}{2},\sqrt{\frac{N}{\pi}}a$ and $\sqrt{\frac{N}{\pi}}\frac{3}{2}a$, respectively. The main panel shows the transverse displacement and the inset shows the longitudinal displacement. The semiclassical result (purple) is also presented.}
   \label{figS3}
\end{figure}

To be specific, we double and triple the real-space standard deviation $\sigma_r$ and calculate the corresponding dynamics with the results shown in Fig.~\ref{figS3}. One can clearly see that the time evolution of $\langle\hat x\rangle$ approaches the semiclassical limit (purple curve) as the real-space spread increases. In fact, this is also true for $\langle\hat y\rangle$ (inset in Fig.~\ref{figS3}), which stays unchanged after the initial recoil and takes on the value $2a$ in the semiclassical limit which can be obtained by calculating Eq.~(\ref{avgrs}) in the $\sigma_k\to0$ limit.

\subsection{Influence of the phase of $l_\uparrow$}
While the Berry curvature 
\begin{equation}
\Omega_{xy}(\boldsymbol{k})=-\frac{8|l_\uparrow|^2\cos k_xa\cos k_ya}{(|l_\uparrow|^2+4\sin^2k_xa+4\sin^2k_ya)^2}
\label{BC}
\end{equation}
depends only on the magnitude of $l_\uparrow$ and so does the semiclassical dynamics, its phase does affect the wave-packet dynamics through the initial recoil. This is clear from Eqs.~(\ref{expk}) and (\ref{avgrs}), especially from the factors ${\rm Re}(l_{\boldsymbol{k}\uparrow}l^*_{\boldsymbol{k}\downarrow})$ and ${\rm Im}(l_{\boldsymbol{k}\uparrow}l^*_{\boldsymbol{k}\downarrow})$. 
This Berry curvature (\ref{BC}) gives a zero Chern number for arbitrary $l_\uparrow$.

\begin{figure}
\begin{center}
        \includegraphics[width=8cm, clip]{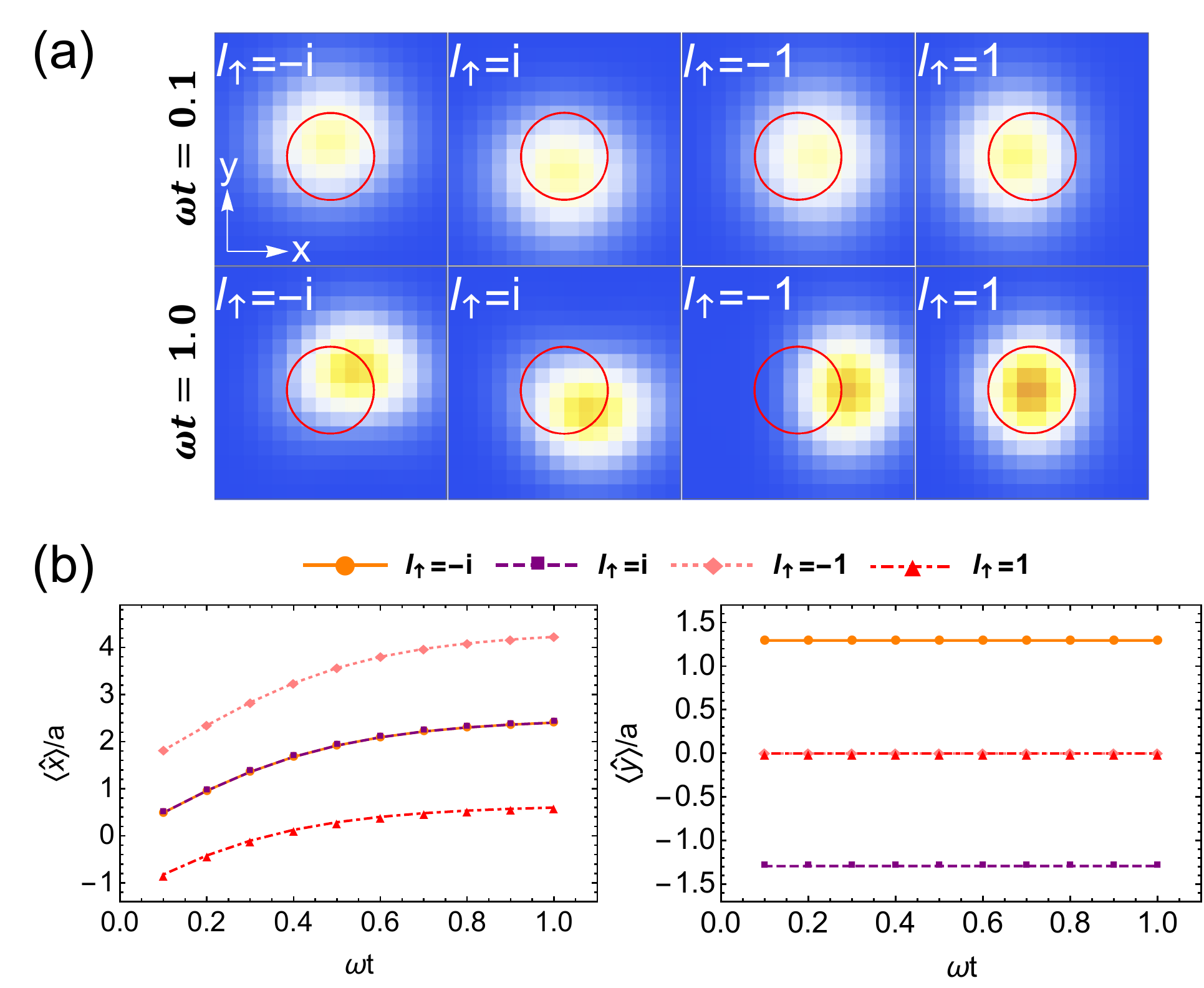}\\
      \end{center}
   \caption{(a) Snapshots of the wave packets for $l_\uparrow=\pm1,\pm i$ at time $\omega t=0.1$ and $\omega t=1.0$. While the initial recoils occur along four different directions, all of the wave packets move in the positive $x$-axis. The red circles mark the initial positions of the wave packets. (b) Time evolutions of the center of mass for $l_\uparrow=-i$ (orange solid), $i$ (purple dashed), $-1$ (pink dotted) and $1$ (red dash-dotted). Here $\langle\hat{x}\rangle$ and $\langle\hat{y}\rangle$ denote the transverse and logitudinal displacements, respectively. Note that the orange and purple (pink and red) curves  overlap in the left (right) panel.}
   \label{figS4}
\end{figure}

We perform numerical calculations for four different phases $l_\uparrow=\pm1,\pm i$, with the results shown in Fig.~\ref{figS4}. One can see that the initial recoils occur along four different directions, despite the fact that due to the ZHE the subsequent motions occur all along the $x$-axis. Furthermore, the $\langle\hat x\rangle$ -- $t$ curves for $l_\uparrow=1$ and $-1$ are not parallel to each other due to the recoils in the BZ [see Eq.~(\ref{expk})] along the positive and negative $k_y$-axis, respectively.

\section{Formal analytic solution}
By defining $\psi_-(\boldsymbol{k},t)\equiv\frac{Na}{2\pi}\langle\boldsymbol{k}-|\psi(t)\rangle$, the projected Schr\"odinger equation reads
\begin{equation}
i\hbar\partial_t\psi_-(\boldsymbol{k},t)=-\boldsymbol{F}\cdot[i\nabla_{\boldsymbol{k}}+\boldsymbol{\mathcal{A}}(\boldsymbol{k})]\psi_-(\boldsymbol{k},t),
\label{PSE}
\end{equation}
which is invariant under the following gauge transformation stemming from the phase degree of freedom of $|\boldsymbol{k}-\rangle$:
\begin{equation}
\begin{split}
\psi_-(\boldsymbol{k},t)\to e^{i\alpha(\boldsymbol{k})}\psi_-(\boldsymbol{k},t),\\
\boldsymbol{\mathcal{A}}(\boldsymbol{k})\to\boldsymbol{\mathcal{A}}(\boldsymbol{k})+\nabla_{\boldsymbol{k}}\alpha(\boldsymbol{k}).
\end{split}
\end{equation}
Suppose that the initial condition is $\psi_-(\boldsymbol{k},0)=\varphi(\boldsymbol{k})$. Without loss of generality, we can write down the trial solution as 
\begin{equation}
\psi_-(\boldsymbol{k},t)=e^{i\theta(\boldsymbol{k},t)}\varphi(\boldsymbol{k}-\boldsymbol{F}t/\hbar),
\label{trial}
\end{equation}
where $\theta(\boldsymbol{k},t)$ is not necessarily real and satisfies $\theta(\boldsymbol{k},0)=0$. Substituting Eq.~(\ref{trial}) into Eq.~(\ref{PSE}), we obtain
\begin{equation}
(\hbar\partial_t+\boldsymbol{F}\cdot\nabla_{\boldsymbol{k}})\theta(\boldsymbol{k},t)=\boldsymbol{F}\cdot\boldsymbol{\mathcal{A}}(\boldsymbol{k}).
\label{thetaE}
\end{equation}
Subjected to the initial condition $\theta(\boldsymbol{k},0)=0$, the solution to Eq.~(\ref{thetaE}) can be expressed as
\begin{equation}
\theta(\boldsymbol{k},t)=\int^t_0dt'\frac{\boldsymbol{F}}{\hbar}\cdot\boldsymbol{\mathcal{A}}\left(\boldsymbol{k}+\frac{\boldsymbol{F}(t'-t)}{\hbar}\right),
\end{equation}
which turns out to be always real, so that $|\psi_-(\boldsymbol{k},t)|^2=|\varphi(\boldsymbol{k}-\boldsymbol{F}t/\hbar)|^2$. This implies that in momentum space, the wave packet undergoes ballistic motion with the density profile unchanged. This fact validates the use of Gaussian smoothing (note that $\varphi(\boldsymbol{k})$ is approximately Gaussian) in the main text.

\section{Further information on the retroreflection}

In this section, we provide a detailed argument and extensive numerical results to support the universality of the retroreflection phenomenon.

\subsection{Detailed argument on the universality}
\label{arg}
We argue in details that, driven by a potential gradient $\boldsymbol{F}$, a wave packet colliding with a boundary by an anomalous velocity alone will be retroreflected. The argument consists of two parts: (i) The Berry curvature landscape features both positive (``hill") and negative (``valley") values whenever $\hat L_{\boldsymbol{r}}$ is short-ranged; (ii) The wave packet always diffuses from valleys to hills or vice versa via collision with a boundary. 

In the main text, we have briefly mentioned (i) and two relevant references \cite{Tang2014} and \cite{Diehl2015}. 
Reference \cite{Tang2014} proves the theorem 
that in an \emph{isolated} lattice system with  shorted-ranged hoping exactly flat band has zero Chern number, 
while Ref.~\cite{Diehl2015}, which claims that for a quadratic Lindblad equation with short-ranged jump operators the \emph{unique mixed steady state} is topologically trivial. However, neither of these results is directly applicable to our system. Fortunately, the QZ subspace is a Hilbert subspace after all, so is a band. Furthermore, the QZ subspace can actually be regarded as the flat band of an isolated lattice system described by $\hat H_{p}=E\sum_{\boldsymbol{r}}\hat L^\dag_{\boldsymbol{r}}\hat L_{\boldsymbol{r}}$ ($E$ is a nonzero $c$-number), which is short-ranged if and only if $\hat L_{\boldsymbol{r}}$ is short-ranged. Therefore, the theorem in Ref.~\cite{Tang2014} can indirectly be applied to our system to validate the statement (i). 

To understand (ii), we first note that the anomalous velocity $\dot{\boldsymbol{r}}$ cannot be orthogonal to the normal vector $\boldsymbol{n}$ of the boundary to be collided with. 
Without the loss of generality, we assume that the wave packet mainly crosses valleys before the collision. At an early stage of the collision, $\dot{\boldsymbol{r}}$ is directed towards the boundary and the wave packet is compressed in the $\boldsymbol{n}$ direction in real space, and thus diffuses in the same direction in the BZ due to the uncertainty relation. At the same time, the wave packet continues to move in the BZ at a constant velocity $\boldsymbol{F}/\hbar$, which is orthogonal to $\dot{\boldsymbol{r}}=-\boldsymbol{\Omega}(\boldsymbol{k})\times\boldsymbol{F}/\hbar$. Hence, $\boldsymbol{F}/\hbar$ and $\boldsymbol{n}$ are not parallel and span the entire BZ, implying that the wave packet can spread to any point in the BZ. Once reaching the hills, the wave packet leaves the boundary and the compression in real space stops, so does the diffusion in the BZ. Therefore, the wave packet eventually enters the orbit that mainly crosses the hills.

\subsection{Influence of an irregular boundary}
\begin{figure}
\begin{center}
        \includegraphics[width=8cm, clip]{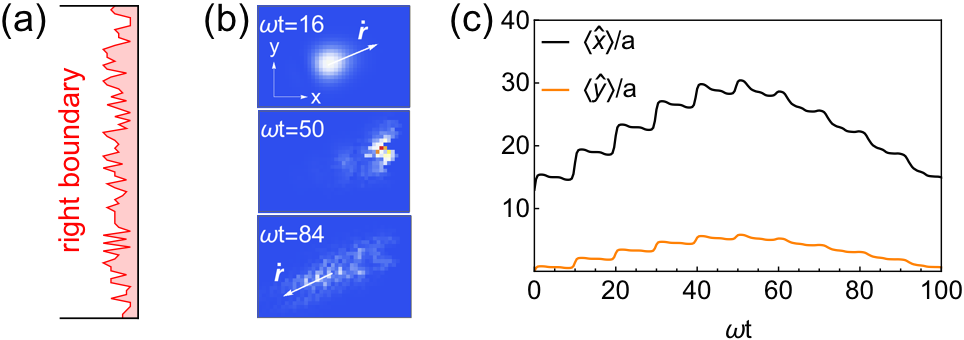}\\
      \end{center}
   \caption{(a) Irregular right boundary (red) created by randomly imposing strong local one-body loss on several sites near the original flat boundary (black). (b) Real-space density profiles of the wave packet before ($\omega t=16$), at ($\omega t=50$) and after ($\omega t=84$) the collision with the irregular right boundary. (c) Real-space dynamics of the center of mass, which clearly shows that retroreflection occurs even at the irregular boundary.}
   \label{figSr4}
\end{figure}

Since the anomalous velocity has only two possible directions, we may expect that the retroreflection is robust against an imperfectly flat boundary. To confirm such robustness, we consider a worst-case situation in which the boundary is so irregular even at the spatial scale of the wave packet that the normal vector $\boldsymbol{n}$ cannot be defined (see Fig.~\ref{figSr4} (a)). Such a boundary can be created also by making use of the QZ effect -- we randomly pick out a narrow region (light red in Fig.~\ref{figSr4} (a)), where we impose strong local one-body loss on each site. The bulk property of the system and the initial state, which is a Gaussian packet located at $\boldsymbol{r}=(13,0)a$ with the spin state being $|\downarrow\rangle$, are the same as those in the main text. 

 While the boundary is highly irregular, the wave packet turns out to be again retroreflected (see Fig.~\ref{figSr4} (b) and (c)) \cite{MVSM}, although the profile after the retroreflection differs significantly from that for a flat boundary (see Fig.~4 in the main text). We note that the argument in Sec.~\ref{arg} is still applicable after a slight modification that the diffusion from valleys to hills occurs in many different directions in the absence of a well-defined $\boldsymbol{n}$.

\subsection{Influence of $\boldsymbol{F}$}

\begin{figure}
\begin{center}
        \includegraphics[width=8cm, clip]{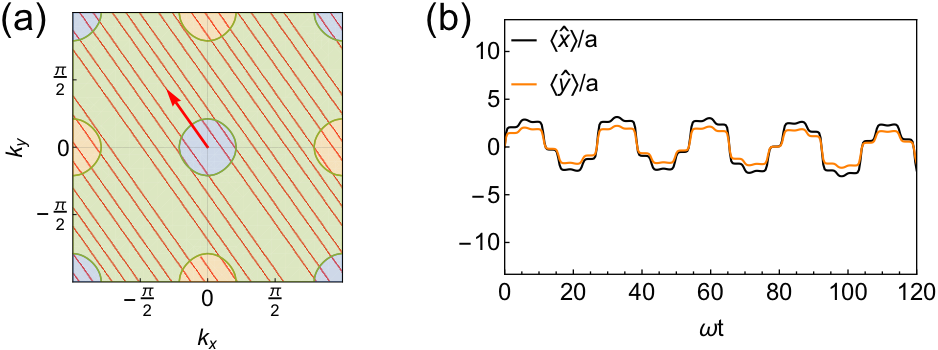}\\
      \end{center}
   \caption{(a) Trajectory (red line) in the BZ generated by a constant velocity $\dot{\boldsymbol{k}}=\frac{F}{\sqrt{3}\hbar}(-1,\sqrt{2})$ (red arrow). The blue and orange regions correspond to the valleys and hills, respectively. (b) Real-space dynamics of the center of mass.}
   \label{figSr5}
\end{figure}

In the main text, the ratio $F_y/F_x$($=-3$) is set to be rational so that the trajectory in the BZ forms a closed orbit with well-defined winding numbers ($1$ for $k_x$ and $3$ for $k_y$ directions), invalidating the ergodicity. The ergodicity, together with zero Chern number, is expected to prohibit the wave packet from approaching a boundary. Note that the center-of-mass motion of the wave packet is governed by the local Berry curvature. However, the wave packet undergoes positive and negative Berry curvatures by roughly the same amount due to the ergodicity and zero Chern number, and hence it cannot move far away from its original position. 

As a bad example, we consider the wave-packet dynamics for $\boldsymbol{F}=\frac{F}{\sqrt{3}}(-1,\sqrt{2})$ in the same model in the main text. While $\dot{\boldsymbol{k}}=\boldsymbol{F}/\hbar$ stays constant, the trajectory in the BZ shows clear ergodicity and crosses the valleys (blue) and hills (orange) almost equally (see Fig.~\ref{figSr5} (a)), leading to \emph{nonperiodic} oscillation in the real space (see Fig.~\ref{figSr5} (b)). After a long time $\omega t\sim10^2$, the wave packet is still near the original position and fails to approach a boundary.

\begin{figure}
\begin{center}
        \includegraphics[width=8cm, clip]{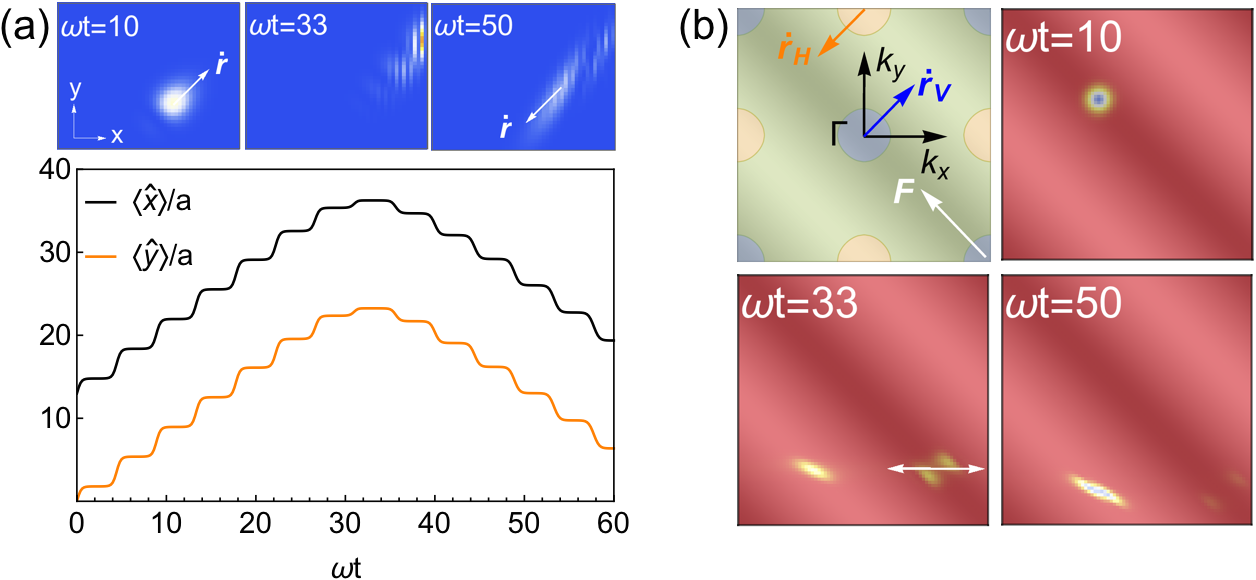}\\
      \end{center}
   \caption{(a) Long-time dynamics of the center of mass for the model in the main text but with a different potential gradient $\boldsymbol{F}=\frac{F}{\sqrt{2}}(-1,1)$. Three panels on the top left show the real-space density profiles of the wave packet before ($\omega t=10$), at ($\omega t=33$) and after ($\omega t=50$) the collision with the right boundary, clearly showing the retroreflection. (b) Hills (orange, $\Omega_{xy}>1$) and valleys (blue, $\Omega_{xy}<-1$) in the BZ and the momentum-space density profiles of the wave packet before, at and after retroreflection. Dark (light) stripes, which are parallel to $\boldsymbol{F}$, refer to the orbits in which the anomalous motion in real space approaches (leaves) the right boundary.}
   \label{figSr6}
\end{figure}

Instead, if we choose $\boldsymbol{F}=\frac{F}{\sqrt{2}}(-1,1)$, the trajectory in the BZ again forms a closed orbit but with different winding numbers ($1$ for both $k_x$ and $k_y$ directions). In this case, the wave packet approaches and collides with the right boundary, and the retroreflection occurs (see Fig.~\ref{figSr6}). The wave packet in the BZ does not fragment and the step length of motion stays unchanged after the collision, since the winding number along the $k_y$ direction is $1$. Such a dynamics differs significantly from that in the main text, where the winding number along the $k_y$ direction is $3$ and thus the wave packet in the BZ fragments into three pieces, leading to a reduction of step length of motion to one-third after the collision.

\subsection{Influence of $\hat L_{\boldsymbol{r}}$}
\begin{figure}
\begin{center}
        \includegraphics[width=8cm, clip]{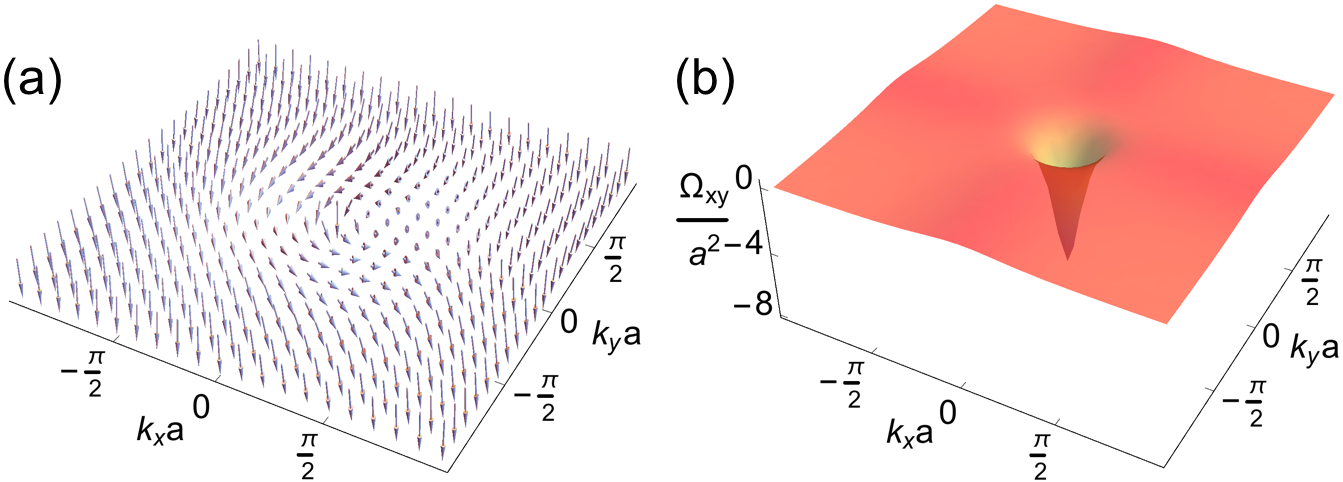}\\
      \end{center}
   \caption{(a) Spin texture and (b) Berry curvature $\Omega_{xy}(\boldsymbol{k})$ of the lower band (QZ subspace) emerging from the engineered dissipation (\ref{LE}) corresponding to the jump operators in Eq.~(\ref{JOp}) with $l'_{\uparrow}=-5$.}
   \label{figSr1}
\end{figure}

To further support the universality of the retroreflection phenomenon, we consider yet another model which features a special 
Berry curvature landscape. The jump operator is chosen to be
\begin{equation}
\begin{split}
\hat L'_{\boldsymbol{r}}&=l'_{\uparrow}\hat c_{\boldsymbol{r}\uparrow}+\hat c_{\boldsymbol{r}+\boldsymbol{a}_x,\uparrow}+\hat c_{\boldsymbol{r}+\boldsymbol{a}_y,\uparrow}+\hat c_{\boldsymbol{r}-\boldsymbol{a}_x,\uparrow}+\hat c_{\boldsymbol{r}-\boldsymbol{a}_y,\uparrow}\\
&+\hat c_{\boldsymbol{r}+\boldsymbol{a}_x,\downarrow}+i\hat c_{\boldsymbol{r}+\boldsymbol{a}_y,\downarrow}-\hat c_{\boldsymbol{r}-\boldsymbol{a}_x,\downarrow}-i\hat c_{\boldsymbol{r}-\boldsymbol{a}_y,\downarrow},
\end{split}
\label{JOp}
\end{equation}
according to which one can check that the Berry curvature of the emergent band reads
\begin{equation}
\begin{split}
&\Omega'_{xy}(\boldsymbol{k})=-2(\cos k_x+\cos k_y+l'_{\uparrow})\\
&\times\frac{(\cos k_x+\cos k_y+l'_{\uparrow}\cos k_x\cos k_y)}{[\sin^2k_x+\sin^2k_y+(\cos k_x+\cos k_y+l'_{\uparrow})^2]^2},
\end{split}
\end{equation}
as plotted in Fig.~\ref{figSr1} (b). Unlike $\Omega_{xy}(\boldsymbol{k})$ in the main text, $\Omega'_{xy}(\boldsymbol{k})$ has only one deep valley, where the spin texture becomes the most twisted (see Fig.~\ref{figSr1} (a)). 

\begin{figure}
\begin{center}
        \includegraphics[width=8cm, clip]{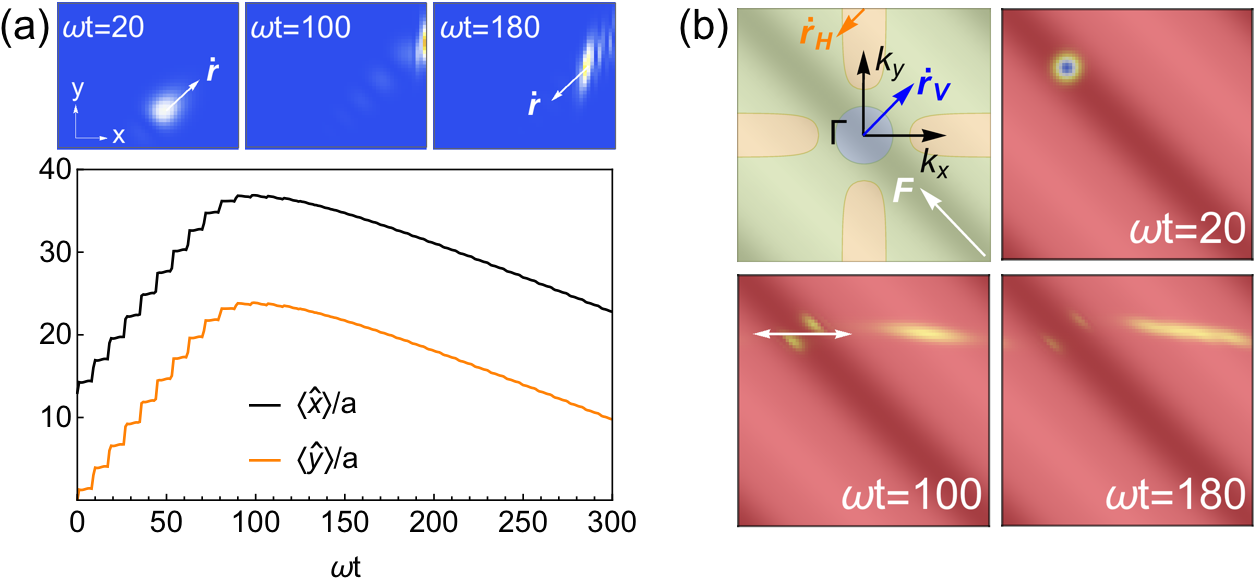}\\
      \end{center}
   \caption{(a) Long-time dynamics of the center of mass for the model (\ref{JOp}). Three panels on the top left show the real-space density profiles of the wave packet before ($\omega t=20$), at ($\omega t=100$) and after ($\omega t=180$) the collision with the right boundary, indicating the retroreflection. (b) Hills (orange, $\Omega_{xy}>0.2$) and valleys (blue, $\Omega_{xy}<-0.2$) in the BZ and the momentum-space density profiles of the wave packet before, at and after the retroreflection. Dark (light) stripes, which are parallel to $\boldsymbol{F}=\frac{F}{\sqrt{2}}(-1,1)$, refer to the orbits in which the anomalous motion in real space approaches (leaves) the right boundary.}
   \label{figSr2}
\end{figure}

We numerically calculate the long-time dynamics for this new model (\ref{JOp}) in response to a potential gradient $\boldsymbol{F}=\frac{F}{\sqrt{2}}(-1,1)$. The initial state is the same as that in the main text. Similar to Fig.~\ref{figSr6} (and Fig.~4 in the main text), we observe stepwise motion and retroreflection (see Fig.~\ref{figSr2}). Nevertheless, the profile of the retroreflected wave packet differs clearly from that in Fig.~\ref{figSr6} due to the very different 
Berry curvature landscape (see Fig.~\ref{figSr1}). 
While the wave packet also evolves into a single branch in the BZ instead of fragmenting, it is more diffusive along the $k_x$ direction in Fig.~\ref{figSr2} than in Fig.~\ref{figSr6}, because the hills in this model cover a relatively large area. Such a specific Berry curvature landscape also smears out the stepwise feature of the motion after the retroreflection, because there are always some components of the wave packet on the hills.

\subsection{Comparison with ordinary reflection}

\begin{figure}
\begin{center}
        \includegraphics[width=8cm, clip]{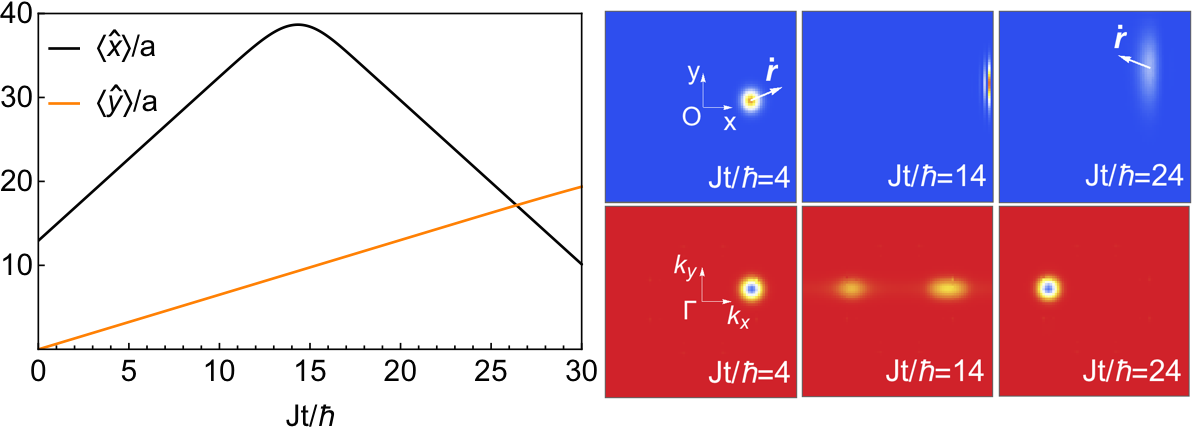}\\
      \end{center}
   \caption{Real-space dynamics of the center of mass of a wave packet with initial velocity $\boldsymbol{v}=\frac{2Ja}{3\hbar}(3,1)$ and reflected by a boundary. Three pairs of the wave-packet profiles in both real space and the BZ before ($Jt/\hbar=4$), at ($Jt/\hbar=14$) and after ($Jt/\hbar=24$) the collision with the right boundary are shown in the right panels.}
   \label{figSr3}
\end{figure}

Finally, it is worthwhile to compare retroreflection with ordinary reflection. 
For the latter case, we only have to consider the simplest tight binding model $\hat H_0=-J\sum_{\boldsymbol{r}}(\hat c^\dag_{\boldsymbol{r}+\boldsymbol{a}_x}\hat c_{\boldsymbol{r}}+\hat c^\dag_{\boldsymbol{r}+\boldsymbol{a}_y}\hat c_{\boldsymbol{r}}+{\rm H.c.})$, with $J$ being the nearest neighbor hopping. Unlike the anomalous velocity, adding a potential gradient to $\hat H_0$ leads to the Bloch oscillations and cannot make the wave packet approach a boundary. Instead, we prepare a Gaussian packet with a finite velocity $\boldsymbol{v}=\frac{2Ja}{3\hbar}(3,1)$, whose $v_x$-component is maximized. The subsequent unitary dynamics without a potential gradient clearly shows ordinary reflection (see Fig.~\ref{figSr3}). The anisotropic diffusion is consistent with analytical calculations, which give $\sigma^2_\mu(t)-\sigma^2_\mu(0)\simeq(\frac{Ja^2t}{\hbar\sigma_\mu(0)})^2[1-(\frac{\hbar v_\mu}{2Ja})^2]$ ($\mu=x,y$), with $\sigma^2_{x,y}(t)$ being the spread in the $x,y$-direction at time $t$ and $\sigma^2_\mu(0)=\frac{Na^2}{4\pi}$.

\section{Details of the experimental implementation}
\subsection{$\Lambda$ energy-level configuration of $^{87}$Rb in an optical lattice}
\begin{figure}
\begin{center}
        \includegraphics[width=6cm, clip]{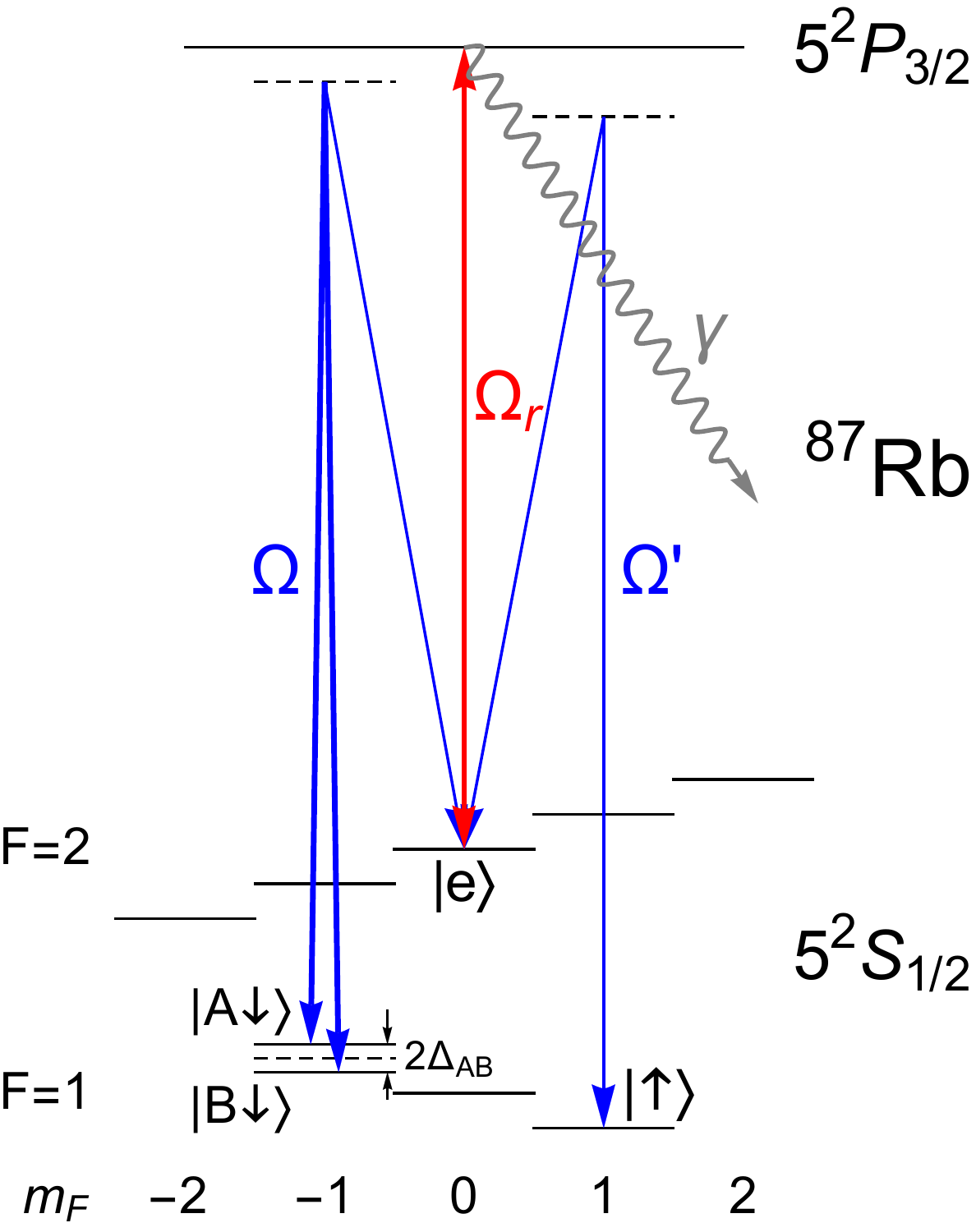}\\
      \end{center}
   \caption{Schematic illustration of the $\Lambda$ energy levels constructed from $|\downarrow\rangle=|F=1,m_F=-1\rangle$, $|\uparrow\rangle=|F=1,m_F=1\rangle$ and $|e\rangle=|F=2,m_F=0\rangle$ within the $\rm 5^2S_{1/2}$ manifold of $^{87}$Rb. Here $|\downarrow\rangle$ and $|\uparrow\rangle$ are coupled to $|e\rangle$ via Raman transitions. The Raman laser that excites $|\downarrow\rangle$ is slightly dichromatic due to the fact that the $|\downarrow\rangle\leftrightarrow|e\rangle$ transition occurs between the nearest-neighbor sites with an energy imbalance $\Delta_{AB}$. 
   Besides $|e\rangle$ is resonantly coupled to the electronic excited state $5^2P_{3/2}$, which undergoes spontaneous emission to modes outside of the $5^2S_{1/2},\;F=1$ manifold at rate $\gamma$.}
   \label{figS5}
\end{figure}

As mentioned in the main text, the fundamental ingredients to engineer a nonlocal loss are an on-site loss and a fine-tuned Rabi coupling.  For $^{87}$Rb atoms in an optical lattice, the sign of the polarizability of the ground state $\rm 5^2S_{1/2}$ and that of the first excited state $\rm 5^2P_{3/2}$ are opposite, implying that they are trapped at two different sets of lattice sites at which the laser intensity reaches the minimum and the maximum. 
If $|e\rangle$ is chosen to be $\rm 5^2P_{3/2}$, the loss will not be on-site with respect to $\rm 5^2S_{1/2}$. 
Therefore, we construct the $\Lambda$ energy-level configuration 
from the ground state manifold $\rm 5^2S_{1/2}$ such as $|\downarrow\rangle=|F=1,m_F=-1\rangle$, $|\uparrow\rangle=|F=1,m_F=1\rangle$ and $|e\rangle=|F=2,m_F=0\rangle$ (see Fig.~\ref{figS5}). Here $|e\rangle$ can be made unstable via coupling to the $\rm 5^2P_{3/2}$ manifold 
by a resonant laser with Rabi frequency $\Omega_r$ which can be tuned to control the effective loss rate  \cite{Smerzi2014}.

While $|e\rangle$ can be directly coupled to $|\uparrow\rangle$ ($|\downarrow\rangle$) by a circularly polarized microwave, we use the Raman coupling because the target Rabi frequency should change over the spatial scale of the optical lattice, which is significantly smaller than the wavelength of the microwave. The details of the Raman coupling are given in the next subsection. The staggered structure of the square lattice can be created by superimposing a superlattice potential
\begin{equation}
V_S(\boldsymbol{x})=V_{S0}\left[\sin^2\left(k_S\frac{x+y}{\sqrt{2}}\right)+\sin^2\left(k_S\frac{x-y}{\sqrt{2}}\right)\right]
\end{equation}
onto the normal square lattice potential 
\begin{equation}
V_L(\boldsymbol{x})=V_{L0}(\sin^2k_Lx+\sin^2k_Ly),
\end{equation}
where $k_S=\frac{2\pi}{\lambda_S}$ and $k_L=\frac{2\pi}{\lambda_L}=\frac{\pi}{a}$. It is clear that $V_S(\boldsymbol{x})$ can be created from two orthogonal standing waves of the same wavelength $\lambda_S$ along the directions $\frac{\boldsymbol{e}_x+\boldsymbol{e}_y}{\sqrt{2}}$ and $\frac{\boldsymbol{e}_x-\boldsymbol{e}_y}{\sqrt{2}}$. By choosing $\lambda_S=\sqrt{2}\lambda_L$, we obtain
\begin{equation}
V_S(\boldsymbol{x})=V_{S0}(1-\cos k_Lx\cos k_Ly),
\end{equation}
implying $V_S(\boldsymbol{r})=V_{S0}[1-(-)^{\boldsymbol{r}}]$ with $(-)^{\boldsymbol{r}}\equiv(-)^{m+n}$ if $\boldsymbol{r}=(m,n)a$ ($m,n\in\mathbb{Z}$), and an energy imbalance $\Delta_{AB}\simeq2V_{S0}$ between the two sublattices. We define the sublattice with higher (lower) on-site energy as $A$ ($B$) (see Fig.~\ref{figS5} and also Fig.~4 in the main text).

\subsection{Raman laser configuration}
The second important ingredient, i.e., engineering a structured Rabi frequency configuration, has been extensively investigated in the context of artificial gauge fields for neutral atoms \cite{Dalibard2011,Goldman2014}, where the Raman coupling is widely used. Generally, after the rotating-wave approximation, the effective Hamiltonian in the rotating frame for two Raman lasers $\boldsymbol{E}_{\omega_+}e^{-i(\omega+\delta\omega)t}$ and $\boldsymbol{E}_{\omega_-}e^{-i\omega t}$ is given by $\hat H_R=\Omega_z\hat J_z + \Omega_-\hat J_+ + \Omega_+\hat J_-$ \cite{Goldman2014}, where $\Omega^*_+=\Omega_-$ and 
\begin{equation}
\Omega_\pm\propto i(\boldsymbol{E}^*_{\omega_\pm}\times\boldsymbol{E}_{\omega_\mp})\cdot(\boldsymbol{e}_x\pm i\boldsymbol{e}_y).
\label{RamanRabi}
\end{equation}

\begin{figure}
\begin{center}
        \includegraphics[width=8cm, clip]{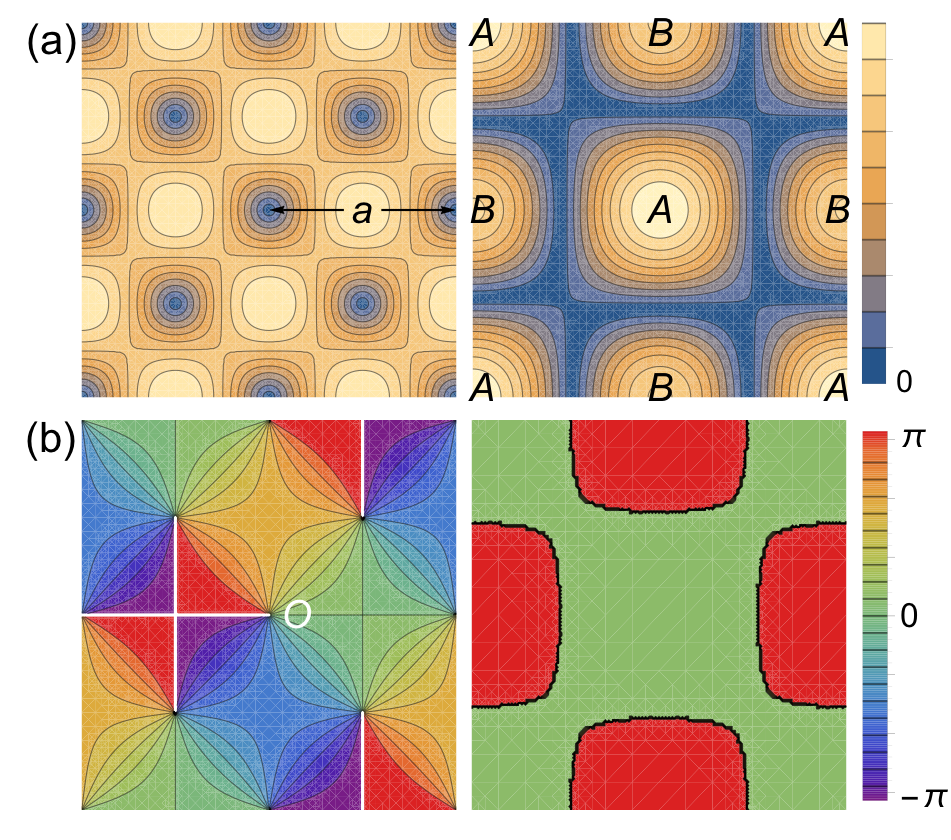}\\
      \end{center}
   \caption{(a) Magnitude $|\Omega(\boldsymbol{x})|$ (not to scale, upper right gauge) and (b) phase ${\rm Arg}\Omega(\boldsymbol{x})$ (lower right gauge) of the Rabi frequency for the $|\downarrow\rangle\leftrightarrow|e\rangle$ (left half column) and $|\uparrow\rangle\leftrightarrow|e\rangle$ (right column) transitions. For $|\downarrow\rangle\leftrightarrow|e\rangle$, the $p$-wave symmetry of the nearest-neighbor hopping manifests itself as the $2\pi$ phase change of $\tilde\Omega_-(\boldsymbol{x})$ around $O$. For $|\uparrow\rangle\leftrightarrow|e\rangle$, the difference in the on-site coupling strength between the sublattices $A$ and $B$ is compensated by the asymmetric spatial pattern of $\tilde\Omega'_+(\boldsymbol{x})$.}
   \label{figS6}
\end{figure}

In the absence of a superlattice, to imprint short-range $p$-wave symmetry for the $|\downarrow\rangle\leftrightarrow|e\rangle$ transition, we apply two sets of counterpropagating, circularly polarized lasers along the $x$- and $y$-axes
\begin{equation}
\begin{split}
\boldsymbol{E}_{\omega_+}\propto \frac{\boldsymbol{e}_y+i\boldsymbol{e}_z}{\sqrt{2}}e^{ik_Rx}+\frac{\boldsymbol{e}_y-i\boldsymbol{e}_z}{\sqrt{2}}e^{-ik_Rx},\\
\boldsymbol{E}_{\omega_-}\propto \frac{\boldsymbol{e}_x+i\boldsymbol{e}_z}{\sqrt{2}}e^{ik_Ry}+\frac{\boldsymbol{e}_x-i\boldsymbol{e}_z}{\sqrt{2}}e^{-ik_Ry},
\end{split}
\label{cpcp}
\end{equation}
where $k_R=k_L$ (strictly speaking, with a tiny relative difference around $10^{-5}$) and $\delta\omega>0$ resonant with the transition between $|\downarrow\rangle$ and $|e\rangle$ (or with a detuning much smaller than the Rabi frequency). Substituting Eq.~(\ref{cpcp}) into Eq.~(\ref{RamanRabi}) yields
\begin{equation}
\tilde\Omega_-(\boldsymbol{x})=g(\sin k_Rx\cos k_Ry+i\cos k_Rx\sin k_Ry).
\label{pRabi}
\end{equation}
The magnitude and phase patterns of $\tilde{\Omega}_-(\mathbf{x})$ are plotted in the left column of Fig.~\ref{figS6}. Here $\tilde\Omega_-(\boldsymbol{x})=2\langle F=2,m_F=0|J_+|F=1,m_F=-1\rangle\Omega_-(\boldsymbol{x})$ which means that there is only a difference in the constant factor between $\tilde\Omega_-(\boldsymbol{x})$ and $\Omega_-(\boldsymbol{x})$ 
($\tilde\Omega'(\boldsymbol{x})$ (\ref{osRabi}) can be understood similarly). We note that the $p$-wave symmetry completely suppresses the on-site $|\downarrow\rangle\leftrightarrow|e\rangle$ transition, leaving the nearest-neighbor coupling dominant. These results hold true in the presence of a superlattice, except for that the frequency difference of the Raman lasers should be modified as $\delta\omega\pm\Delta_{AB}$. Namely, we have to replace $\boldsymbol{E}_{\omega_+}$ with $\boldsymbol{E}^{(1)}_{\omega_+}$ and $\boldsymbol{E}^{(2)}_{\omega_+}$, which have the same strength, polarization and propagation direction but a tiny frequency difference $2\Delta_{AB}$ (see Fig.~\ref{figS5}). We note that in this case if there is a small on-site $|\downarrow\rangle\leftrightarrow|e\rangle$ Rabi frequency due to some imperfection, such a transition will be suppressed by a relatively large $\Delta_{AB}$.

To realize the on-site $|\downarrow\rangle\leftrightarrow|e\rangle$ transition, we apply another linearly polarized standing wave along the $x$-axis
\begin{equation}
\boldsymbol{E}'_{\omega_+}\propto \boldsymbol{e}_z\cos k_Rx,
\label{sw}
\end{equation}
which is much weaker than $\boldsymbol{E}_{\omega_+}$ and the frequency difference $\delta\omega'<0$ from $\boldsymbol{E}_{\omega_-}$ is resonant with the $|\downarrow\rangle\leftrightarrow|e\rangle$ transition. Such a choice (\ref{sw}) leads to the Rabi frequency pattern
\begin{equation}
\tilde\Omega'_+(\boldsymbol{x})=g'\cos k_Rx\cos k_Ry.
\label{osRabi}
\end{equation}
Here $g'$ should be much smaller than $g$, so as to make the Rabi frequency ($\Omega$) for the nearest-neighbor hopping $|\downarrow\rangle\leftrightarrow|e\rangle$ comparable with that ($\Omega'$) for the on-site $|\uparrow\rangle\leftrightarrow|e\rangle$ transition. Due to the superlattice structure, Eq.~(\ref{osRabi}) alone leads to an asymmetry between sublattices $A$ and $B$ and thus to a deviation from the target dynamics. Nevertheless, such an asymmetry can easily be eliminated by a uniform Rabi frequency achieved by a weak microwave resonant with the on-site $|\downarrow\rangle\leftrightarrow|e\rangle$ transition. The spatial pattern of the Rabi frequency is visualized in the right column of Fig.~\ref{figS6}, where the compensation from the microwave has been taken into account.

In practice, we can 
rotate the Raman laser beams out of the $xy$-plane by a small angle $\theta$. Then we have
\begin{equation}
\begin{split}
\boldsymbol{E}^\theta_{\omega_+}&\propto \frac{\boldsymbol{e}_y+i(\boldsymbol{e}_z\cos\theta-\boldsymbol{e}_x\sin\theta)}{\sqrt{2}}e^{ik_R(z\sin\theta+x\cos\theta)}\\
&+\frac{\boldsymbol{e}_y-i(\boldsymbol{e}_z\cos\theta+\boldsymbol{e}_x\sin\theta)}{\sqrt{2}}e^{ik_R(z\sin\theta-x\cos\theta)},\\
\boldsymbol{E}^\theta_{\omega_-}&\propto \frac{\boldsymbol{e}_x+i(\boldsymbol{e}_z\cos\theta-\boldsymbol{e}_y\sin\theta)}{\sqrt{2}}e^{ik_R(z\sin\theta+y\cos\theta)}\\
&+\frac{\boldsymbol{e}_x-i(\boldsymbol{e}_z\cos\theta+\boldsymbol{e}_y\sin\theta)}{\sqrt{2}}e^{ik_R(z\sin\theta-y\cos\theta)},\\
\boldsymbol{E}^\theta_{\omega'_+}&\propto(\boldsymbol{e}_z\cos\theta-\boldsymbol{e}_x\sin\theta)e^{ik_R(z\sin\theta+x\cos\theta)}\\
&+(\boldsymbol{e}_z\cos\theta+\boldsymbol{e}_x\sin\theta)e^{ik_R(z\sin\theta-x\cos\theta)}.
\end{split}
\end{equation}
After lengthy but straightforward calculations, we obtain the corresponding Rabi frequencies
\begin{equation}
\begin{split}
\tilde\Omega^\theta_-(\boldsymbol{x})&=(1-\sin\theta)\cos\theta\tilde\Omega_-(\boldsymbol{x}\cos\theta),\\
\tilde\Omega'^\theta_+(\boldsymbol{x})&=(1-\sin\theta)\cos\theta\tilde\Omega'_+(\boldsymbol{x}\cos\theta)-\Delta\tilde\Omega'^\theta_+(\boldsymbol{x}),
\end{split}
\end{equation}
where $\Delta\tilde\Omega'^\theta_+(\boldsymbol{x})=ig'\sin\theta\sin (k_Rx\cos\theta)\sin (k_Ry\cos\theta)$, $\Omega_-(\boldsymbol{x})$ and $\Omega'_+(\boldsymbol{x})$ are given in Eqs.~(\ref{pRabi}) and (\ref{osRabi}). 
Except for the additional term $\Delta\tilde\Omega'^\theta_+(\boldsymbol{x})$ in $\tilde\Omega'^\theta_+(\boldsymbol{x})$, which clearly preserves the checkerboard pattern due to $\Delta\tilde\Omega'^\theta_+(\boldsymbol{x}+\frac{\lambda_R}{2\cos\theta}\boldsymbol{e}_{x,y})=-\Delta\tilde\Omega'^\theta_+(\boldsymbol{x})$ and can be compensated by an additional Raman laser $\boldsymbol{E}^{-\theta}_{\omega'_+}$, the only differences lie in the fact that $g$ (or $g'$) is multiplied by a factor $\cos\theta(1-\sin\theta)$ and $k_R$ should be replaced by $k_R\cos\theta$, where the relation $\lambda_R=\lambda_L\cos\theta$ must be satisfied to make the spatial pattern of the Rabi coupling coincide with that of the optical lattice. In this sense, $\lambda_R$($<\lambda_L$) can be chosen independent of $\lambda_L$.

\subsection{Adiabatic elimination}
Let us derive an effective open quantum system dynamics involving only $|\downarrow\rangle$ and $|\uparrow\rangle$ by adiabatically eliminating other degrees of freedom. We start from the following Lindblad equation
\begin{equation}
\dot{\hat\rho}=-\frac{i}{\hbar}[\hat H,\hat\rho]+\kappa\int d^2\boldsymbol{x}\mathcal{D}[\hat\psi_e(\boldsymbol{x})]\hat\rho,
\label{cLE}
\end{equation}
where $\kappa=\frac{|\Omega_r|^2}{\gamma}$ with $\gamma^{-1}$ being the lifetime of $\rm 5^2P_{3/2}$, provided that $\gamma\gg|\Omega_r|$, 
and $\hat H=\int d^2\boldsymbol{x}\hat{\mathcal{H}}(\boldsymbol{x})$ is the two-dimensional integral of the Hamiltonian density
\begin{equation}
\begin{split}
\hat{\mathcal{H}}(\boldsymbol{x})=\hat{\mathcal{H}}_0(\boldsymbol{x})&+\frac{1}{2}[\tilde\Omega_-(\boldsymbol{x})\hat\psi^\dag_e(\boldsymbol{x})\hat\psi_\downarrow(\boldsymbol{x})+{\rm H.c.}]\\
&+\frac{1}{2}[\tilde\Omega'_+(\boldsymbol{x})\hat\psi^\dag_e(\boldsymbol{x})\hat\psi_\uparrow(\boldsymbol{x})+{\rm H.c.}].
\end{split}
\end{equation}
Here $\hat{\mathcal{H}}_0(\boldsymbol{x})$ consists of the kinetic term, the optical lattice potential and the interaction. We then perform the tight-binding approximation by expanding the field operator as $\hat\psi_\nu(\boldsymbol{x})\simeq\sum_{\boldsymbol{r}}w_{X(\boldsymbol{r})}(\boldsymbol{x}-\boldsymbol{r})\hat c_{\boldsymbol{r}\nu}$ ($\nu=\uparrow,\downarrow,e$), where $w_{X(\boldsymbol{r})}(\boldsymbol{x})$ ($X(\boldsymbol{r})\in\{A,B\}$) is the Wannier function. In terms of $\hat c_{\boldsymbol{r}\nu}$, Eq.~(\ref{cLE}) is approximated as
\begin{equation}
\dot{\hat\rho}=-\frac{i}{2}[\sum_{\boldsymbol{r},\boldsymbol{r}',\sigma}\Omega_{\boldsymbol{r},\boldsymbol{r}',\sigma}\hat c^\dag_{\boldsymbol{r}e}\hat c_{\boldsymbol{r}'\sigma}+{\rm H.c.},\hat\rho]+\kappa\sum_{\boldsymbol{r}}\mathcal{D}[\hat c_{\boldsymbol{r}e}]\hat\rho,
\label{tightLE}
\end{equation}
provided that the Rabi coupling and dissipation are dominant and the intrinsic hoping and the interaction stemming from $\mathcal{H}_0(\boldsymbol{x})$ are negligible. Here $\Omega_{\boldsymbol{r},\boldsymbol{r}',\sigma}=(-)^{\boldsymbol{r}}\Omega_{\boldsymbol{r}'-\boldsymbol{r},\sigma}$, $\Omega_{\boldsymbol{r}\downarrow}=\Omega(\delta_{\boldsymbol{r},\boldsymbol{a}_x}+i\delta_{\boldsymbol{r},\boldsymbol{a}_y}-\delta_{\boldsymbol{r},-\boldsymbol{a}_x}-i\delta_{\boldsymbol{r},-\boldsymbol{a}_y})$, and $\Omega_{\boldsymbol{r}\uparrow}=\Omega'\delta_{\boldsymbol{r},\boldsymbol{0}}$, where the coefficients $\Omega$ and $\Omega'$ are given for the simplest case of $\theta=0$ as
\begin{equation}
\begin{split}
\Omega&=g\int d^2\boldsymbol{x} w^*_B(\boldsymbol{x}-\boldsymbol{a}_x)w_A(\boldsymbol{x})\sin k_Rx\cos k_Ry,\\
\Omega'&=\frac{g'}{2}\int d^2\boldsymbol{x} (|w_B(\boldsymbol{x})|^2+|w_A(\boldsymbol{x})|^2)\cos k_Rx\cos k_Ry.
\end{split}
\end{equation}


The operator $\hat c_{\boldsymbol{r}e}$ in Eq.~(\ref{tightLE}) can be adiabatically eliminated by using the general formula derived in Ref.~\cite{Sorensen2012}: For a Lindblad equation $\dot{\hat\rho}=-\frac{i}{\hbar}[\hat H,\hat\rho]+\sum_j\mathcal{D}[\hat L_j]\hat\rho$ whose jump operators satisfy $\hat P_g\hat L_j\hat P_e=\hat L_j$, with $P_g$ and $P_e$ being the projection operators onto the ground and the excited subspaces, the effective dynamics involving only the ground-state manifold in the large decay limit is described by 
\begin{equation}
\dot{\hat\rho}=-\frac{i}{\hbar}[\hat H',\hat\rho]+\sum_j\mathcal{D}[\hat L'_j]\hat\rho, 
\end{equation}
where
\begin{equation}
\begin{split}
\hat H'&=\hat H_g-\frac{1}{2}\hat V_-[\hat H^{-1}_{\rm NH}+(\hat H^{-1}_{\rm NH})^\dag]\hat V_+,\\
\hat L'_j&=\hat L_j\hat H^{-1}_{\rm NH}\hat V_+,
\end{split}
\end{equation}
with $\hat H_{g,e}\equiv\hat P_{g,e}\hat H\hat P_{g,e}$, $\hat V_+=\hat V^\dag_-\equiv\hat P_e\hat H\hat P_g$, and $\hat H_{\rm NH}\equiv\hat H_e-\frac{i\hbar}{2}\sum_j\hat L^\dag_j\hat L_j$. For Eq.~(\ref{tightLE}) and by neglecting the corrections of the order of magnitude no more than $O(\frac{|\Omega|^2}{\kappa^2})$ in the $\kappa\gg|\Omega|$ limit ($|\Omega|$ is the typical magnitude of $\Omega_{\boldsymbol{r},\boldsymbol{r}',\sigma}$), we have $\hat P_g=\sum_{\hat c_{\boldsymbol{r}e}|\Psi\rangle=0,\forall\boldsymbol{r}}|\Psi\rangle\langle\Psi|$, $\hat L_j=\hat c_{\boldsymbol{r}e}$, $\hat H_e\simeq\hat H_g=0$, $\hat V_+=\hat V^\dag_-=\frac{\hbar}{2}\sum_{\boldsymbol{r},\boldsymbol{r}',\sigma}\Omega_{\boldsymbol{r},\boldsymbol{r}',\sigma}\hat c^\dag_{\boldsymbol{r}e}\hat c_{\boldsymbol{r}'\sigma}\hat P_g$ and $\hat H_{\rm NH}\simeq-\frac{i\hbar\kappa}{2}\hat N_e\simeq-\hat H^\dag_{\rm NH}$ ($\hat N_e=\sum_{\boldsymbol{r}}\hat c^\dag_{\boldsymbol{r}e}\hat c_{\boldsymbol{r}e}$), so that 
\begin{equation}
\begin{split}
\hat H'&\simeq0,\\
\hat L'_{\boldsymbol{r}}&\simeq\frac{i}{\kappa}\hat c_{\boldsymbol{r}e}\hat N^{-1}_e\sum_{\boldsymbol{r}'',\boldsymbol{r}',\sigma}\Omega_{\boldsymbol{r}'',\boldsymbol{r}',\sigma}\hat c^\dag_{\boldsymbol{r}''e}\hat c_{\boldsymbol{r}'\sigma}\hat P_g\\
&=\sum_{\boldsymbol{r}'',\boldsymbol{r}',\sigma}\frac{i\Omega_{\boldsymbol{r}'',\boldsymbol{r}',\sigma}}{\kappa}\hat c_{\boldsymbol{r}e}\hat c^\dag_{\boldsymbol{r}''e}\hat c_{\boldsymbol{r}'\sigma}\hat P_g\\
&=\sum_{\boldsymbol{r}',\sigma}\frac{i\Omega_{\boldsymbol{r},\boldsymbol{r}',\sigma}}{\kappa}\hat c_{\boldsymbol{r}'\sigma}\hat P_g.
\end{split}
\label{effHL}
\end{equation}
Here we have used the fact that for an arbitrary $|\Psi\rangle$ which satisfies $\hat P_g|\Psi\rangle=|\Psi\rangle$, $\hat c^\dag_{\boldsymbol{r}e}|\Psi\rangle$ is an eigenstate of $\hat N_e$ with eigenvalue $1$. Equation (\ref{effHL}) implies the following effective dynamics:
\begin{equation}
\dot{\hat\rho}=\Gamma\sum_{\boldsymbol{r}}\mathcal{D}[\hat L_{\boldsymbol{r}}]\hat\rho,
\label{FLE}
\end{equation}
where $\Gamma=\frac{|\Omega|^2}{\kappa}$ and $\hat L_{\boldsymbol{r}}=\sum_{\boldsymbol{r}',\sigma}\Omega_{\boldsymbol{r},\boldsymbol{r}',\sigma}\hat c_{\boldsymbol{r}'\sigma}/\Omega$. We note that Eq.~(\ref{FLE}) is even more general than Eq.~(1) in the main text, since the translational invariance has not been assumed at this stage.

Recalling the fact that $\mathcal{D}[\alpha\hat L]\hat\rho=|\alpha|^2\mathcal{D}[\hat L]\hat\rho$ and $\Omega_{\boldsymbol{r},\boldsymbol{r}',\sigma}=(-)^{\boldsymbol{r}}\Omega_{\boldsymbol{r}'-\boldsymbol{r},\sigma}$ for our specific engineering, Eq.~(\ref{FLE}) stays unchanged if the jump operator is replaced by $\hat L_{\boldsymbol{r}}=\sum_{\boldsymbol{r}',\sigma}l_{\boldsymbol{r}'-\boldsymbol{r},\sigma}\hat c_{\boldsymbol{r}'\sigma}$ with $l_{\boldsymbol{r}'-\boldsymbol{r},\sigma}=\Omega_{\boldsymbol{r}'-\boldsymbol{r},\sigma}/\Omega$, which indicates the same translational symmetry as the lattice. 

\subsection{Energy and time scales}

\begin{figure}
\begin{center}
        \includegraphics[width=8cm, clip]{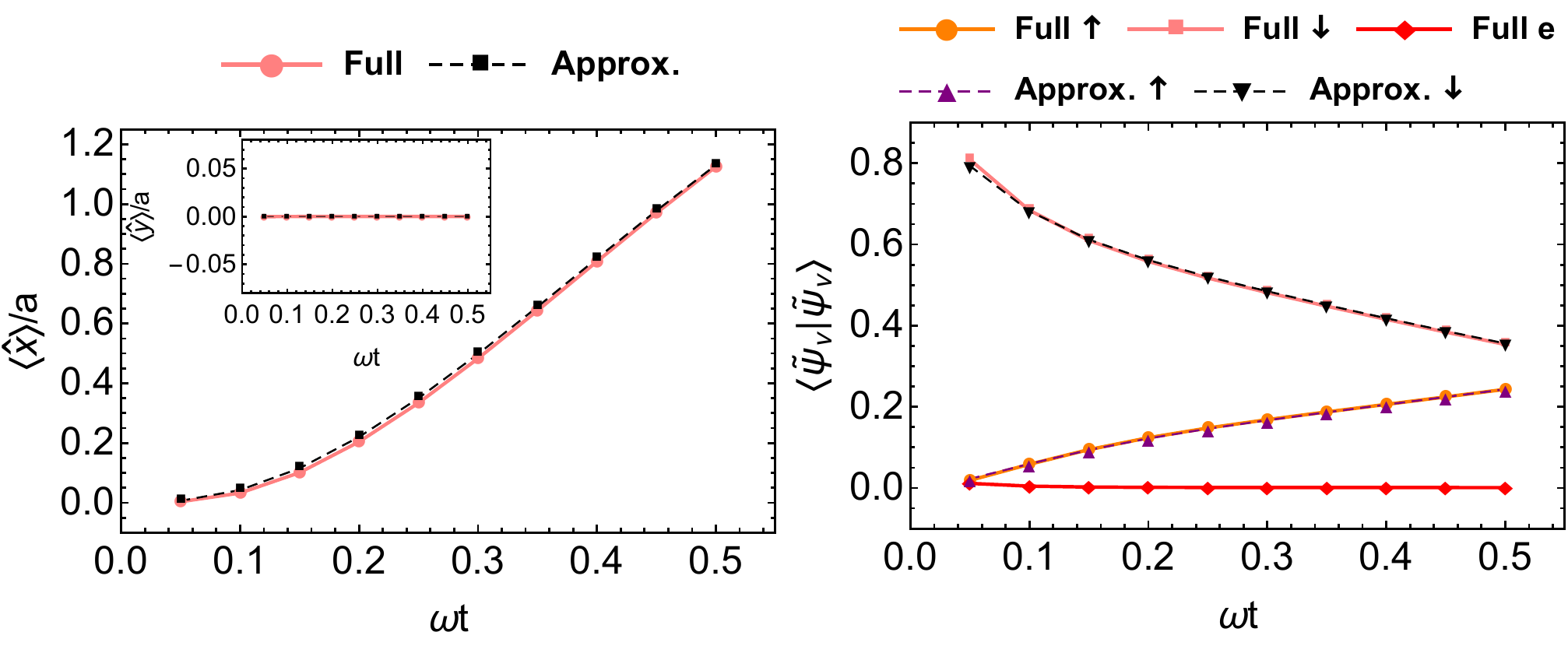}\\
      \end{center}
   \caption{Comparison between the full open quantum system dynamics (\ref{tightLE}) including the internal state $|e\rangle$ (solid curves, marked by ``Full"), and the effective dynamics (\ref{FLE}) after adiabatically eliminating $|e\rangle$ (dashed curves, marked by ``Approx."). We can see good agreement for both the real-space dynamics (left panel) and the internal state populations (right panel), where the $|e\rangle$ component stays negligible. The parameters are chosen to be $\Omega=0.2\kappa$, $\omega=0.1\Gamma$ and $N=40$, and the initial internal state is chosen to be $|\downarrow\rangle$.}
   \label{figS7}
\end{figure}

For the short-time dynamics, the time scale of the experiment is $\omega^{-1}$, which should be made much smaller than the decoherence time $\tau_{\rm de}$. Besides inhomogeneity of a magnetic field (the corresponding $\tau_{\rm de}$ can be made as long as several hundreds of miliseconds), which induces a Zeeman splitting about several MHz that is clearly separated from the energy scale of the hyperfine splitting and that of the optical potentials, photon scattering may be the dominant decoherence mechanism. In particular, the laser, which resonantly couples $|F=2,m_F=0\rangle$ to the $5^2P_{3/2}$ excited state, may be the main source of decoherence for the $F=1$ manifold, since the detuning from the $F=1$ state is fixed at around $\delta\sim 2\pi\times6$ GHz, namely the splitting between the $F=1$ and $F=2$ hyperfine states. Note that $\delta\gg\gamma\sim2\pi\times6$ MHz, $\tau^{-1}_{\rm de}$ can thus be estimated as $\frac{|\Omega_r|^2}{\delta^2}\gamma=\frac{\gamma^2}{\delta^2}\kappa\sim10^{-6}\kappa$. On the other hand, we must have $\omega\ll\Gamma=\frac{|\Omega|^2}{\kappa}\ll|\Omega|\ll\kappa$, which allows $\omega\sim 4\times10^{-3}\kappa\gg\tau^{-1}_{\rm de}$ if we choose $\omega=0.1\Gamma$ and $\Omega=0.2\kappa$. The decoherence caused by the laser that couples $F=2$ to $5^2P_{3/2}$ turns out to be negligible in this case. 
Typical experimental parameters are $\kappa=10$ kHz, $\Omega=2$ kHz, $\Gamma=0.4$ kHz and $\omega=40$ Hz. 
We have numerically confirmed that the adiabatic elimination gives a good approximation in this case (see Fig.~\ref{figS7}). While $\Gamma=10\omega$ is still far from the quantum Zeno limit, we can observe the onset of the QZ effect as demonstrated in Figs.~\ref{figS1} and \ref{figS2}. Further numerical calculations show a transverse displacement of $0.93a$ at $t=0.4\omega^{-1}$ ($10$ ms for $\omega=40$ Hz) for the initial spin state $|\downarrow\rangle$, which is considerably larger than the displacement of $0.56a$ for the initial spin state $\frac{1}{\sqrt{2}}(|\downarrow\rangle+|\uparrow\rangle)$. This can be understood from the fact that less time is needed to project a Gaussian wave packet onto the QZ subspace from the spin state $|\downarrow\rangle$ than from the state $\frac{1}{\sqrt{2}}(|\downarrow\rangle+|\uparrow\rangle)$, since the spin state of the lower band is close to $|\downarrow\rangle$ near $\boldsymbol{k}=\boldsymbol{0}$ (see Fig.~2 (a) in the main text).

For other photon scattering processes such as the Raman couplings and the optical lattice potential, the decoherence rates are roughly $\frac{\Omega_R}{\delta_R}\gamma$ and $\frac{V_{L0}}{\delta_L}\gamma$ ($\frac{V_{S0}}{\delta_S}\gamma$), respectively, where $\Omega_R$ is the typical magnitude of $\tilde\Omega(\boldsymbol{x})$, $\delta_R$ and $\delta_L$ ($\delta_S$) are the detunings from the $5^2S_{1/2}\to5^2P_{3/2}$ transition. These quantities can be made even smaller than $10^{-8}\gamma<1$ Hz by using a large detuning ranging from several to over one hundred THz. 
To this end, we use lasers with a short wavelength such as $\lambda_L=532$ nm to create the optical lattice, so that $\delta_R\lesssim\delta_L\sim2\pi\times180$ THz and $\delta_S\sim2\pi\times14$ THz. In this case, the recoil energy reads $E_{r}=\frac{4\pi^2\hbar^2}{2m\lambda^2_L}\sim\hbar\times50$ kHz. For $V_{L0}=5E_r$, we obtain the bare tunneling $J\simeq0.06E_r\sim\hbar\times3$ kHz \cite{Bloch2008}, which sets an upper bound $2\times0.6J/\hbar\sim4$ kHz on $\Omega$ \cite{Goldman2014}, and can be suppressed by an energy imbalance $\Delta_{AB}\gtrsim0.3E_r$ between the sublattices. The on-site Rabi coupling of $|\downarrow\rangle$ to the higher bands is expected to be no more than $\Delta_{AB}$, 
which is negligible since the band gap near $\boldsymbol{k}=\boldsymbol{0}$ is larger than $4E_r$.

On the other hand, monitoring the long-time dynamics seems to be challenging due to the decoherence. Nevertheless, if we can prepare the wave packet extremely close to the boundary of a box potential \cite{Hadzibabic2013}, the signature of retroreflection may still be observed in a relatively short time like tens of $\omega^{-1}$. The initial state can be prepared by a harmonic trap and a set of $|\Delta m|=2$ Raman lasers 
that couples $|\uparrow\rangle$ to $|\downarrow\rangle$, so that the wave packet is spatially Gaussian and in a definite spin state such as $|\downarrow\rangle$.

\end{document}